\documentclass[aps, preprint, nofootinbib,preprintnumbers,eqsecnum]{revtex4}

\usepackage[dvips]{color}
\usepackage[normalem]{ulem}
\usepackage{amsmath}
\usepackage{amssymb}
\usepackage{amscd}
\usepackage{enumerate}
\usepackage{amsfonts}
\usepackage{epsfig}
\usepackage{yfonts}

\usepackage{dsfont}

\usepackage{graphicx}
\usepackage{bm}

\definecolor{davecolor}{rgb}{0.95,  0.5,  0.2}

\def\({\left(}
\def\){\right)}
\def\[{\left[}
\def\]{\right]}
\def\<{\langle}
\def\>{\rangle}

\def\tr{\mathop{\rm tr}}

\newcommand\half{{\ensuremath{\frac{1}{2}}}}

\newcommand{\be}{\begin{equation}}
\newcommand{\ee}{\end{equation}}
\newcommand{\bea}{\begin{eqnarray}}
\newcommand{\eea}{\end{eqnarray}}
\newcommand{\bwt}{\begin{widetext}}
\newcommand{\ewt}{\end{widetext}}

\newcommand{\bi}{\begin{itemize}}
\newcommand{\ei}{\end{itemize}}
\newcommand{\ben}{\begin{enumerate}}
\newcommand{\een}{\end{enumerate}}
\newcommand{\bca}{\begin{cases}}
\newcommand{\eca}{\end{cases}}
\newcommand{\bln}{\begin{align}}
\newcommand{\eln}{\end{align}}
\newcommand{\bst}{\begin{split}}
\newcommand{\est}{\end{split}}

\def\g{\lambda}

\begin{document}

\title{Bulk emergence and the RG flow of Entanglement Entropy}
\begin{abstract}
We further develop  perturbative methods used to calculate entanglement entropy (EE) 
away from an interacting CFT fixed point. 
 At second order we find certain universal terms in the renormalized EE which were predicted previously from holography and which we find hold universally for relevant deformations of any CFT in any dimension. 
We use both replica methods and direct methods to calculate the EE and in both cases find a non-local  integral expression involving the CFT two point function. 
We show that this integral expression can be written as a local integral over a higher dimensional \emph{bulk} modular hamiltonian in an emergent $AdS$ space-time. 
This bulk modular hamiltonian is associated to an emergent scalar field dual to the deforming
operator.  We generalize to arbitrary spatially dependent couplings where a linearized metric 
emerges naturally as a way of efficiently encoding the field theory entanglement: by demanding  that Einstein's equations coupled to the bulk scalar field are satisfied, we show that EE can be calculated as the area of this metric.
Not only does this show a direct emergence of a higher dimensional gravitational theory from any CFT, it allows for effective evaluation of the the integrals required to calculate EE perturbativly. 
Our results can also be interpreted as relating the non-locality of the modular hamiltonian for a spherical region in non-CFTs and the non-locality of the holographic bulk to boundary map.

\end{abstract}
\author{Thomas Faulkner}
\affiliation{ University of Illinois, Urbana-Champaign}
\email{tomf@illinois.edu}

\maketitle

\section{Introduction}

The calculation of Entanglement Entropy in QFTs turns out to be a rather non-trivial endeavor. 
EE is a non-local observable capable of revealing and quantifying many non-perturbative aspects of QFT \cite{Holzhey:1994we,Calabrese:2004eu,Kitaev:2005dm,Levin:2006zz}. However it's usefulness, as a theoretical tool, at this point in time is limited by our ability to calculate it. 
This is an unfortunate situation. Even more so now that there are many hints that EE holds a key to a new level of understanding for quantum gravity \cite{Swingle:2009bg,VanRaamsdonk:2010pw,Maldacena:2013xja}. 

One  breakthrough, relevant for this work, came via the CHM construction \cite{Casini:2011kv} where the authors gave us efficient tools to calculate EE in Conformal Field Theories (CFTs). 
In this paper we plan to make some modest steps forward by further developing perturbative methods to deform away from CFTs and understand the scale dependence of EE as we do this. We will be limited to conformal perturbation theory in some relevant coupling $\g$ about a UV fixed point and so our results will only apply for small entangling regions compared to the inverse mass scale of the perturbation away from the initial CFT in the UV. 

A basic motivation for studying EE in QFT comes from its utility in quantifying renormalization
group (RG) flows, via
monotonic $c$-functions in two \cite{Casini:2004bw} and three dimensions \cite{Casini:2012ei}. For example, in 3d relativistic QFT the following function defined in terms of the EE of a ball region of radius $R$
turns out to be a monotonically decreasing function of $R$,
\be
\label{Ffunc}
F(R) = R S_{EE}'(R) - S_{EE}(R)
\ee
evaluating to a constant $F_{UV,IR}$ at  the UV and IR fixed points with a value
intrinsic to the fixed point \cite{Casini:2011kv,Jafferis:2010un,Jafferis:2011zi,Closset:2012vg,Klebanov:2011gs}. An interesting observation in \cite{Liu:2012eea}, using a holographic calculation
based on the Ryu-Takayanagi (RT) conjecture \cite{Ryu:2006bv,Ryu:2006ef,Nishioka:2009un},
was the existence of non-analytic terms in the small $R$ limit (UV) related to the dimension $\Delta$
of the deforming operator,\footnote{We are essentially assuming that $\Delta > d/2$ - when this is
not the case other terms which have non-analytic powers of $\g$ appear at \emph{leading} order. These are discussed in \cite{Nishioka:2014kpa}. Since they cannot be seen in perturbation theory we don't really have any hope of finding these terms in this paper.}
\be
\label{checkhong}
\qquad F(R) = F_{UV} - \frac{2\pi^2 (\Delta -3)} {(2 \Delta -7)} \g^2 R^{2(3-\Delta)}  + \ldots
\ee
These terms were found using holography, yet it seems reasonable they should survive
to any QFT even ones without a classical gravitational dual. In particular one may guess they can be seen in second order perturbation theory about the UV fixed point. Indeed one of the results of this paper is to reproduce these terms exactly from a purely  field theoretic calculation. We will then show that our methods can be generalized to non-uniform couplings $\g(x)$ and a more detailed comparison to holography will emerge. The particular result we would like advertise is the statement
that EE in deformed CFTs can be calculated using a classical general relativity problem.
More precisely:

\emph{ EE for ball regions  in any d-dimensional CFT deformed by a non uniform coupling $\g$ of a relevant operator can be determined to second order in the perturbation  by first solving the following classical general relativity problem in one higher dimension,
\begin{align}
\nabla^2 \phi & = \Delta(\Delta -d ) \phi  \qquad
R_{ab} - \frac{1}{2} R g_{ab} - \frac{1}{2} d(d-1) g_{ab} = 8 \pi G_N T^\phi_{ab}  \qquad
\phi \mathop{\rightarrow}_{z\rightarrow 0} \g (x) z^{d- \Delta} + \ldots
\end{align}
with regular boundary conditions in the interior of the emergent space.  
Where we take the metric to be asymptotically AdS in Poincare coordinates with radial coordinate $z$ as $z \rightarrow 0$. 
After solving this problem at  first order in $\g$ for the scalar perturbation and second order for the metric perturbation about $AdS$,  the EE is proportional to the area of a minimal surface
ending on the ball shaped region at $z=0$. }

In this paper the above result will hold for perturbations $\g(x)$ of the Euclidean theory, so
the gravitational problem stated above is in imaginary time, although generalizations to real times are certainly possible.\footnote{As will be discussed later, the notion
of EE for this problem in Euclidean signature is not always well defined, rather we should be talking
about generalized entropy as in \cite{Lewkowycz:2013nqa}.}
This problem is clearly exactly the one we would solve
if we wanted to calculate EE in holographic theories with a classical gravity dual, to second order in $\g$.

Depending on the readers background, this result may either sound obviously wrong or obviously correct. We are clearly more sympathetic to the later viewpoint.
In some sense it is obviously correct because, as will be seen in this paper, perturbed
EE essentially only depends on the CFT two point function for $\mathcal{O}$ and the $T_{\mu\nu}
\mathcal{O} \mathcal{O}$ three point functions. Since these are universal in any CFT,
including holographic theories, the above result is not at all surprising. Of course it is a non-trivial fact
that EE, a highly non-local observable, depends only on this local data (at least in perturbation theory.)
Further, since this calculation will turn out to be non-trivial, we hope that interesting
lessons about AdS/CFT \cite{Maldacena:1997re,Gubser:1998bc,Witten:1998qj}  can be learned, and extensions to higher order in perturbation theory will be fruitful in that they can see the difference between theories with or without holographic duals.

Previous work along these directions can be found in \cite{Rosenhaus:2014woa,Smolkin:2014hba,Rosenhaus:2014nha,Rosenhaus:2014ula,Rosenhaus:2014zza}. These authors studied
perturbative corrections to CFT EE for both relevant deformations and deformations of
the entangling geometry. Comparing to these works for relevant deformations we find new terms
 that prove important for seeing bulk emergence. 

We also note a possible connection to the works of  \cite{Datta:2014ska,Datta:2014uxa,Datta:2014zpa} for 2 dimension CFTs, where similar universality was noted and proved for the EE of CFTs with W algebra symmetries deformed by current operators at second order.

The plan of the paper is as follows,
after setting up background material in Section~\ref{setup} we turn to applying the replica trick \cite{Holzhey:1994we} to the problem at hand. Here we make progress by relying on certain
standard restults from thermal field theory. In Appendix~\ref{app:direct} we use more direct techniques
to study the same problem - without reference to the replica trick (more along the lines of \cite{Rosenhaus:2014woa}).
We find the same ``non-local'' terms in both methods. In the replica trick, it arrises by a subtle analytic continuation away from integer $n$ and in the direct method, it arrises due to the non-commutivity of the perturbation to the reduced density matrix and the unperturbed density matrix.

In Section~\ref{integrals}-\ref{emergence} we set out to explicitly calculate these ``non-local'' integrals in terms of CFT data. This is where the importance of holography emerges. While we did not succeed in calculating these integrals by brute force, we do so using tricks which re-write them as higher dimensional integrals in terms of an effective dual gravitational theory. 
This method was essentially discovered working backwards from the holographic result. We choose to emphasize the forward direction since it clearly demonstrates how the holographic results hold universally for all CFTs.  In Section \ref{emergence} we generalize to arbitrary spatially dependent couplings and show that the holographic description is still universal. We conclude with open questions and many possibilities for future work.

\section{Setup}
\label{setup}

We are interested in EE  in the vacuum of a QFT for a subregion $A$. From the reduced
density matrix for this sub-region we would like to calculate the quantity:
\be
S_{EE} = - {\rm tr} \rho_A \ln \rho_A \,.
\ee
We will always consider $A$ to be a $d-1$ dimensional ball of radius $R$ on a constant
time like slice of the theory. Further for a CFT
this problem was partially solved in \cite{Casini:2011kv} via a conformal mapping. In particular
for the CFT on Euclidean $\mathbb{R}^d$ space there is a conformal map which takes the theory to $\mathbb{S}^1 \times \mathbb{H}_{d-1} $ where the circumference of $\mathbb{S}^1$ is $2\pi$. The EE then
simply becomes a thermal entropy for the CFT living on spatial slices $\mathbb{H}_{d-1}$. 

To setup notation we start by explaining this conformal mapping. We work in the embedding space
formalism, 
which will be very useful later on, especially
when we relate the results to holography.  Consider a point $P$ in $\mathbb{R}^{1,d+1}$ where:
\be
P^m = \left( P^I, P^{II}, P^\mu \right)
\ee
which lies on the upper light cone:
\be
\label{defemb}
P \cdot P \equiv - (P^I)^2 + (P^{II})^2 + P^\mu P^\mu = 0   \qquad P^I > 0
\ee
After identifying $P \equiv \Lambda P$, also known as projectivizing, for $\Lambda \in \mathbb{R}$ we have remaining a $d$ dimensional space for which the conformal group $SO(d+1,1)$ acts naturally. 
We take our CFT to live on this space. We always have the freedom to rescale $P$ and
gauge fixing this freedom results in different conformally related space-times. The two
important ones for us are, flat euclidean $\mathbb{R}^d$ space:\footnote{We have included
some funny factors of $R$ here for convenience later. The theory on $\mathbb{R}^d$ does not know about the Entangling ball of radius $R$, however these rescalings by $R$ have no real effect on the gauge choice. }
\be
\label{flatgauge}
\left.  \left( P^I, P^{II}, P^\mu \right)  \right|_{F} =   \left( \frac{R^2 + x^2}{2R} , \frac{R^2 - x^2}{2R},
x^\mu \right) \qquad x^\mu \in \mathbb{R}^d
\ee 
where we take $\mu = 0, \ldots d-1$, and the theory in hyperbolic slicing:
\be
\label{Hgauge}
\left.  \left( P^I, P^{II}, P^0,  P^m \right)  \right|_{H} =   \left( Y^I , \cos \tau, \sin \tau, Y^m \right)
\qquad \left(Y^I,Y^m\right) \in \mathbb{H}^{d-1}
\ee
where we use embedding space coordinates for $\mathbb{H}^{d-1}$ defined as the locus $ - (Y^I)^2 + Y^m Y^m  = -1$ and $Y^I >0$ for $m=1 \ldots d-1$. The remaining coordinate $\tau$ is that of $\mathbb{S}^1$. 
Note the relation between these two gauge choices defines the conformal map of interest:
\be
\label{conff}
\left. P \right|_H = \Omega \left. P \right|_F \qquad \Omega = R^{-1} \left( Y^I + \cos \tau \right)
\ee
Note that $x^0$ should be thought of as Euclidean time
and that the boundary of region $A$ which lives at $|\vec{x}| = R, x^0 = 0$
maps to the boundary of $\mathbb{H}^{d-1}$, $Y^I \rightarrow \infty$. Similarly the center of $\mathbb{H}^{d-1}$ ($Y^I=1$) at $\tau = \pi$ maps to the point at infinity on $\mathbb{R}^d$,  $x \rightarrow \infty$.

The thermal ensemble on $\mathbb{H}_{d-1}$ is determined by the modular Hamiltonian generated
by the flow lines of the vector field $\partial_\tau$. This is clearly an isometry of projective space given 
by rotations $P \rightarrow M(\theta) \cdot  P: (P^{II} \pm i P^0) \rightarrow e^{\pm i \theta}  (P^{II} \pm i P^0)$.
For a fixed gauge this will correspond to a conformal isometry since the rotation will need to be accompanied by a rescaling in order to stay in that gauge. For example:
\be
\label{modflow}
\left. \left(M(\theta) \cdot P \right) \right|_F =  \frac{\Omega(P)}{\Omega(M(\theta) \cdot P)}  M(\theta)
\cdot \left( P|_F \right) =   \frac{Y^I + \cos \tau }{Y^I + \cos(\tau + \theta)}  M(\theta)
\cdot \left( P|_F \right) 
\ee
Infinitesimally we can then use this to calculate the conformal killing vector on $\mathbb{R}^d$
which is:
\be
\xi_E = \frac{ R^2 + (x^0)^2 - x^i x^i }{2 R} \partial_0 + \frac{x^0 x^i }{R} \partial_i
\ee
where the spatial coordinates are labelled by $i = 1, \ldots d-1$. 

We can give similar mappings and isometries in real times, where we should wick rotate both our original space and the modular flow parameter:
\be
x^0 \rightarrow - i \sigma \qquad \tau \rightarrow - i s  \qquad \xi_E =  i \xi
\ee
The interpretation is now \cite{Casini:2011kv} a map from the domain of dependence of the region $A$, $\mathcal{D}(A)$
to $\mathbb{R} \times \mathbb{H}_{d-1}$ with $\mathbb{R}$ here corresponding to the real time
direction $s$. 

The modular Hamiltonian which generates the $\xi$ flow in the CFT then corresponds to:
\be
H = \int_{A} d \Sigma^\mu \xi^\nu T_{\mu\nu} =  \int_{A} d^{d-1} x \frac{ (R^2 - x^i x^i)}{2R} T_{\sigma\sigma}
\ee
where the region $A$ lies on the time slice $t= x^0 = 0$.
The reduced density matrix is determined  from the modular energy $H$ as a Gibbs thermal state with
temperature $1/2\pi$. This can be argued by noting the periodicity of the flow generated by $H$ in imaginary times \cite{Casini:2011kv}. That is:
\be
\rho_A = e^{ - 2\pi H}/Z
\ee
Calculating the spectrum of $H$ and from this $S_{EE}$ is still a non-trivial task. In AdS/CFT
$S_{EE}$ can be further related to the entropy of a certain hyperbolic black hole which was then
used in \cite{Casini:2011kv} to give a non-trivial confirmation of the Ryu-Takayanagi conjecture. 
Further arguments along these lines relates the  EE to more conventional CFT observables,
in particular the universal cut-off independent terms are:
\be
\label{fanda}
S_{EE}^{\rm univ}(CFT) =  \left\{ \begin{matrix} \log Z(S^{d}) \,, & \quad d \in {\rm odd} \\ -2 (-1)^{d/2} A_d \log R \,, & \quad  d \in {\rm even} \end{matrix} \right.
\ee
where $A_d$ is the a-type trace anomaly in even dimensions\footnote{The coefficient
of the Euler term in the trace anomaly - we  follow the conventions of \cite{Casini:2011kv} here.} and $Z(S^d)$ is the regularized sphere
partition function of the CFT. We will use these quantities, which are however not always known
for a given CFT,  to fix the normalization of our results later. 

We will need various results on manipulating the projective coordinates $P$,
for example integrating over $P$, distance functions and the relation to embedding space
coordinates for AdS etc. These are  discussed in Appendix~\ref{app:maps}.

For the rest of this paper we will
be using conformal perturbation theory for the problem of calculating EE. All our results can be expressed in terms of integrals of correlation functions living on $\mathbb{S}^1 \times \mathbb{H}_{d-1}$. Since this space is conformally flat we can go fairly far with this. For example we know from general CFT considerations that the form of 2 and 3 point functions of conformal primaries on $\mathbb{R}^d$  is fixed up to a finite set of parameters in terms of the dimension of the operators \cite{Osborn:1993cr}.  In this paper we will essentially only need certain 2  and 3 point functions, however generalizations should work for higher point functions.

\section{Replica Trick}
\label{replica}

The replica trick seeks to calculate the EE using the following recipe. First calculate
the Renyi entropies:
\be
S_n = - \frac{1}{n-1} \ln {\rm tr} \rho_A^n
\ee
for integer $n$. These can be formulated as a path integral $Z_n$ on an $n$-sheeted surface defined
by taking $n$ copies of the original theory in Euclidean space and stitching them together along the co-dimension-$1$ regions $A$.  The boundaries of $A$ host co-dimension $2$ conical singularities
of opening angle $2\pi n$ (a conical excess). We then use:
\be
{\rm tr} \rho_A^n = Z_n/(Z_1)^n
\ee

Notably these path integrals can only be defined for integer $n$,
so the final step is to find a ``nice'' analytic continuation away from integer $n$. The definition of ``nice'' is not entirely clear. It must for example deal with  $\sin( \pi n)$ terms which introduce obvious ambiguities. The general prescription is unknown for QFTs,  although applications of
Carlson's theorem have been successful \cite{Casini:2009sr,Headrick:2010zt,Calabrese:2010he,Cardy:2013nua}.
Here we will follow our
noses a little and see where we end up.  We will cross check our results using a more direct calculation given in Appendix~\ref{app:direct}, so the calculation to follow will in some sense serve as a validation of the replica trick to the situation at hand. 
The reason we are interested in the replica trick in the first place stems from its usefulness
in calculating EE in holographic theories as was discussed in the recent proofs \cite{Faulkner:2013yia,Hartman:2013mia,Lewkowycz:2013nqa}
of the RT formulae using the rules of AdS/CFT.
Further discussions of this analytic continuation in $n$ can be found in Appendix~\ref{app:carlson}.

In the presence of the mass deformation the $Z_n$ partition function can be calculated perturbatively
as follows:
\be
Z_n = \int_{\mathcal{M}_n} \mathcal{D} \phi e^{ - S(\phi) - \int \g \mathcal{O}}
= \int_{\mathcal{M}_n} \mathcal{D} \phi e^{ - S(\phi)} \left( 1 - \g \int_a \mathcal{O}(a)
+ \frac{1}{2}\g^2 \int_{a} \int_{b}  \mathcal{O}(a) \mathcal{O}(b) + \ldots \right)
\ee
where $a,b$ are points on $\mathcal{M}_n$.  The first term is generally speaking easy to deal with and is not of interest to us here, so we may as well assume $\left< \mathcal{O} \right>_{\g=0} = 0$.\footnote{This would be true for a theory with a $Z_2$ symmetry taking $\mathcal{O} \rightarrow -\mathcal{O}$ assuming that this symmetry is not spontaneously broken on the replica manifold.}
 We would like to understand how to calculate the second order term.
So we write this:
\be
Z_n = Z_n(\g=0) \left( 1 + \frac{1}{2} \g^2 \int_a \int_b \left< \mathcal{O}(a) \mathcal{O}(b) \right>_n + \ldots
\right)
\ee

We can write a general expression for the two point function by first making a conformal
transformation to $\mathbb{S}^1 \times \mathbb{H}_{d-1}$ (as in Section~\ref{setup} above) where now in the replicated theory we should work at an inverse temperature $\beta = 2 \pi n$ or
in other words identify $ \tau \equiv \tau + 2 \pi n$. 
Once we do this we can easily see that for conformal primaries: 
\be
  \left< \mathcal{O}(a) \mathcal{O}(b) \right>_n
 = \Omega^{\Delta}(a) \Omega^{\Delta}(b)
 G_n( \tau_b - \tau_a; Y_a \cdot Y_b ) 
\ee
where $\Omega$ is the conformal factor for this mapping given in \eqref{conff} and $Y$ are the embedding
coordinates for $\mathbb{H}_{d-1}$ (see \eqref{distconf}). Due to the symmetries of $\mathbb{H}_{d-1}$, $G_n$ can only depend on the geodesic distance between the two points on  $\mathbb{H}_{d-1}$ and this is only a function of $Y_a \cdot Y_b$ (see Appendix~\ref{app:maps} for discussion of distance functions
on $\mathbb{H}_{d-1}$.)  

Here $G_n$ is a thermal Green's function
for the theory defined on hyperbolic space, with an inverse temperate $\beta = 2 \pi n$.
That is we can use the operator formalism to write:
\be
G_n(\tau_b - \tau_a; Y_a, Y_b) = {\rm Tr} \left(  e^{- 2\pi n H}  \mathcal{T} \widehat{\mathcal{O}}(i \tau_a,Y_a)  \widehat{\mathcal{O}}(i \tau_b,Y_b) \right)
\ee
where time evolution is with respect to the modular hamiltonian $  \widehat{\mathcal{O}}(i \tau_a) = e^{- \tau_a H}  \widehat{\mathcal{O}}(0)
e^{\tau_a H}$ and where $\mathcal{T}$ is the Euclidean time-ordering operation. Note that $G_n$ satisfies the usual properties of Euclidean thermal greens functions.
Including for example the KMS condition which, due to the time ordering operation, simply reads:
\be
G_n(\tau  + 2\pi n; Y_a \cdot Y_b) = G_n( \tau; Y_a \cdot Y_b)
\ee
Also true is the reflection property $G_n(\tau) = G_n(-\tau)$ which follows from the time ordering and symmetry under exchange of $Y_a \leftrightarrow Y_b$. 
In general it is hard to calculate $G_n$ for any given CFT.  
We could for example calculate it
using holography for specific dual theories via the hyperbolic black hole construction
of \cite{Hung:2011nu} for general $n$. However it turns out that the explicit form
of $G_n$ is not required. We only need to know $G_n$ close to $n=1$ where we can calculate
it using conformal mappings and CFT data. Of course we have implicitly assumed we can
continue $G_n$ away from integer $n$ and indeed there is no obstruction to doing this (the theory
is well defined on $\mathbb{H}_{d-1}$ for any temperature.) 

The real difficulty comes from 
finding the correct analytic continuation of: \footnote{See Appendix~\ref{app:maps}  for the definition of $dY$ for
integrating over hyperbolic space in embedding coordinates.}
\be
\delta \ln Z_n = \frac{1}{2} \g^2 \int_0^{2\pi n} \hspace{-.5cm} d\tau_a  \int_0^{2\pi n}  \hspace{-.5cm} d\tau_b 
\int_{\mathbb{H}_{d-1}} \hspace{-.5cm} dY_a  \int_{\mathbb{H}_{d-1}} \hspace{-.5cm}  dY_b  \, \,\Omega^{\Delta-d}(a) \Omega^{\Delta-d}(b)
G_n(\tau_b - \tau_a; Y_a \cdot Y_b )
\ee
For example this is complicated by the conformal factors $\Omega$ given in \eqref{conff}
which do not make sense for $n$ not an integer (under the periodic identification of $\tau_{a,b}
\rightarrow \tau_{a,b} + 2\pi n$.) In other words there is a tension between
the periodicity of the thermal greens function $G_n$ and the conformal factors.

\begin{figure}[h!]
\centering
\includegraphics[scale=.7]{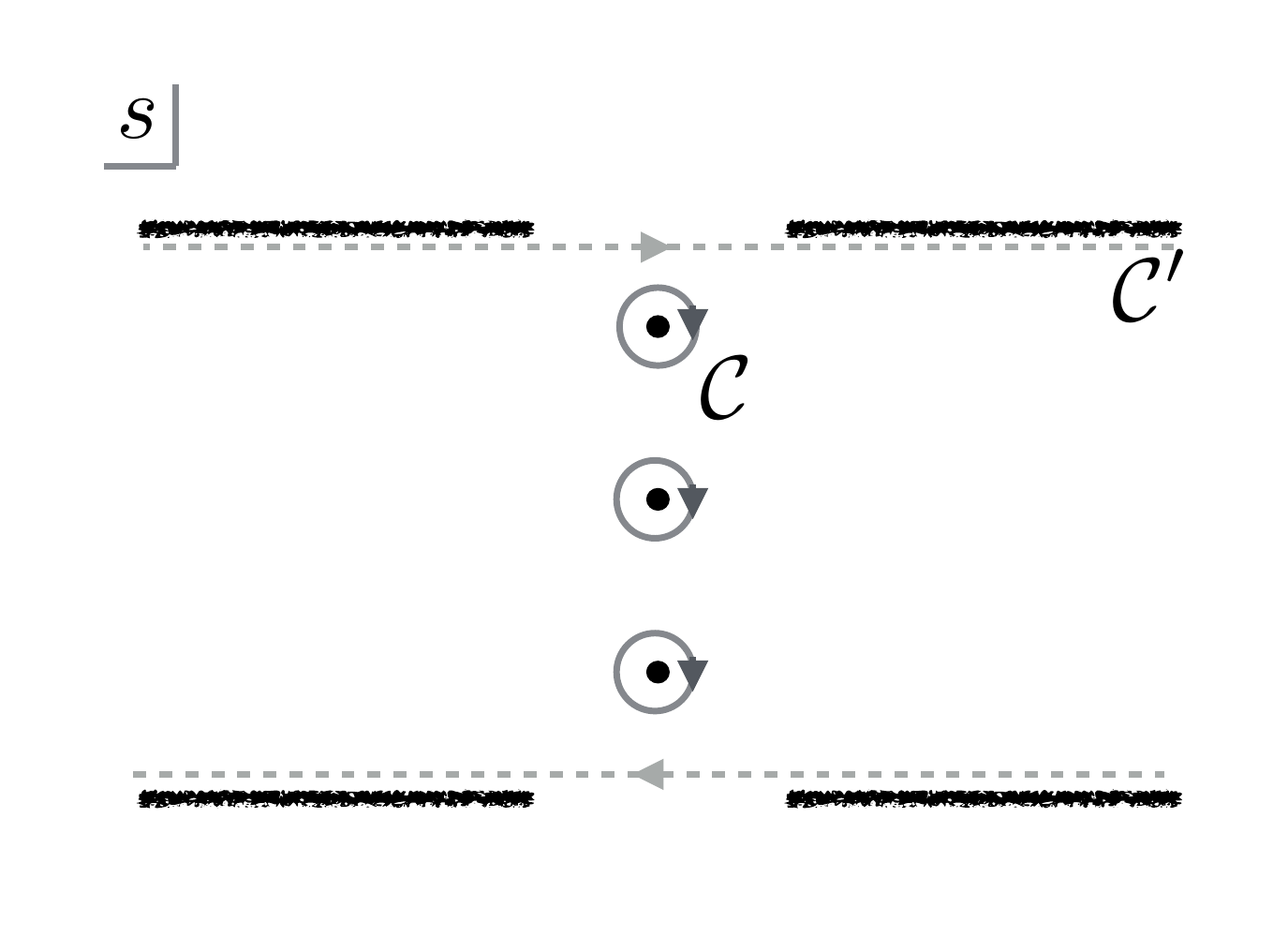}
\vspace{-.7cm}
\caption{ \label{figpoles} Contour integration used to do the sum in \eqref{tosum2} for
the case $n=3$. 
The general branch cut structure of $G_n( - is)$ is shown for any thermal Euclidean greens
functions in the complex time plane (the structure repeats periodically in the imaginary
direction with period $2\pi n$.)
The original contour $\mathcal{C}$ encircles the $n$ poles of $\left(e^{ s- i \tau_{ba}} -1 \right)^{-1}$ between the cuts. We then deform so the
contour $\mathcal{C}'$ lies just above/below the cuts of $G_n$ (dashed lines.)
This method of contour integration for doing replica sums is very similar to the methods
applied to free field theories in \cite{Casini:2009sr}.
}
\end{figure}

To proceed we split the integral over $\tau_{a,b}$ into a sum over the replicas and an integral over $0< \tau_{a,b} < 2\pi$. 
\be
\label{tosum}
\delta \ln Z_n = \int d\mu  \sum_{j_a,j_b =0}^{n-1} G_n(\tau_b - \tau_a + 2\pi (j_a - j_b)  )
\ee
where we have  dropped some superfluous notation and hidden all the integrals in:
\be
\label{dmu}
\int d\mu \ldots  =  \frac{1}{2} \g^2 \int_0^{2\pi } \hspace{-.5cm} d\tau_a  \int_0^{2\pi }  \hspace{-.5cm} d\tau_b 
\int_{\mathbb{H}_{d-1}} \hspace{-.5cm} dY_a  \int_{\mathbb{H}_{d-1}} \hspace{-.5cm}  dY_b  \, \,\Omega^{\Delta-d}(a) \Omega^{\Delta-d}(b) \ldots 
\ee
Note in particular that we have used the periodicity of $\Omega$ for $\tau \rightarrow \tau + 2\pi$.
Next write the double sum as:
\be
\label{dblsum}
\delta \ln Z_n =  \sum_{j_a, j_b = 0}^{n-1} s_n(j_a - j_b)  \qquad s_n(j)
= \int d\mu G_n( \tau_b - \tau_a + 2 \pi j )
\ee
One can show that $s_n(-j) = s_n(j)$ by making use of the reflection condition on $G_n$ as well
as an exchange of the coordinates $a \leftrightarrow b$. Additionally
$s_n(j+n) = s_n(j)$ follows from the KMS condition. Using these two properties we can
reduce the two sums in \eqref{dblsum} to a single sum which can be done using contour integration:
\be
\label{tosum2}
\delta \ln Z_n = n \sum_{j=0}^{n-1} s_n(j) =  n \int d \mu  \sum_{j=0}^{n-1} G_n( \tau_{ba} + 2\pi j )
= \int d\mu \int_{\mathcal{C}} \frac{ds}{2\pi i} \frac{1}{ \left(e^{ s- i \tau_{ba}} -1 \right)}  G_n (- i s)
\ee
where we have written $\tau_{ba} = \tau_b - \tau_a$. The contour
$\mathcal{C}$ encircles the poles of the first term in the integrand located at points $s = i ( \tau_{ba}+  2 \pi k)$ 
for integer $k$ and  which lie between $ 0 < -i s < 2 \pi n$ (note that $-2\pi < \tau_{ba} < 2\pi$ so depending on the sign of
$\tau_{ba}$ we either include the pole with $k=0$ or $k=n$.) We have used the unique analytic continuation of the Euclidean
greens functions $G_n( - i s)$ to the strip between $0 < {\rm Im} s < 2 \pi n $ \cite{lebellac}. 
See Figure~\ref{figpoles}.

We can now deform the contour $\mathcal{C}$ so that it lies just above the real line ${\rm Im} s =0$ and just below ${\rm Im} s = 2 \pi n$. 
\be
\label{twogn}
\delta \ln Z_n = n  \int d\mu \int_{-\infty}^{\infty} \frac{ds}{2 \pi i} \left( \frac{G_n( - i s + \epsilon)}
{ e^{s + i \epsilon- i \tau_{ba}} - 1} - \frac{ G_n( - i s - \epsilon)}{ e^{ s - i \epsilon+ i 2\pi n  - i \tau_{ba}} - 1} \right)  
\ee
We have assumed certain nice behavior for $G_n(i s)$ at large  Re$s$ in order to drop the
vertical integration contours at $s = \pm \infty$. We have checked this for $G_1$ and it
is not hard to argue that it should continue to hold for $n \neq 1$.

The final step is to set $ e^{i 2 \pi n } =1$ in the last term of \eqref{twogn}. It is important to think about this carefully since
this assumes $n$ is an integer. The analytic continuation in $n$ will differ depending on weather
we do this or we don't.\footnote{It seems that this can explain the difference between
the results presented here and the analytic continuation of the Renyi entropies suggested
in Appendix~B of \cite{Lewkowycz:2014jia}. Note however these authors were mostly interested
in the EE of planar regions where the analytic continuation is less subtle because of
the absence of the conformal factors. This should only be true
for deformations that don't break the relatistivic invariance of the underlying theory.
We thank Aitor Lewkowycz for discussion on this. }
The argument for doing this comes from thinking about the function in the complex  $n$ plane. 
Firstly note that $G_n$ is well defined  for real $n>0$, and thus the analytic continuation
of this function to the complex $n$ plane is straightforward (and unnecessary.)\footnote{Of course
$G_n$ could possibly have non-analyticities in $n$ due to thermal phase transitions of
the theory on $\mathbb{H}_{d-1}$ \cite{Belin:2013dva}. As discussed in Appendix~\ref{app:carlson} we don't see this is an obstruction.}   The only issue is
with the denominator term of \eqref{twogn}, this term is non-analytic  at integer spaced poles in the complex $n$ plane. Setting $e^{i 2\pi n} =1$ will remove these poles. We expect the Renyi entropies to have some nice analyticity properties in the complex $n$ plane (at least for ${\rm Re}\, n > 0$), so this is  certainly natural. For further discussion of this see Appendix~\ref{app:carlson}.

If we do set $ e^{i 2\pi n} = 1$ then we get a pleasingly simple result written in terms
of the spectral density:
\be
\label{cont}
\delta \ln Z_n = n \int d\mu \int_{-\infty}^{\infty} \frac{ds}{2\pi} \frac{A_n(s)}
{ \left(e^{s - i \tau_{ba}} -1\right)}
\ee
We claim this is the correct analytic continuation. The spectral density is defined with respect to the CFT living on $\mathbb{H}^{d-1}$ at inverse temperature $2\pi n$:
\be
A_n(s; Y_a, Y_b) = -i {\rm Tr} \left[ e^{ - 2\pi n H} \widehat{\mathcal{O}}_a(0) \widehat{\mathcal{O}}_b(s + i \epsilon)   \right]+i {\rm Tr} \left[ e^{ - 2\pi n H} \widehat{\mathcal{O}}_b(s - i \epsilon)  \widehat{\mathcal{O}}_a(0)   \right]
\ee
where we have introduced the shorthand:  $\widehat{\mathcal{O}}_b(s) \equiv  \widehat{\mathcal{O}}(s,Y_b)$. 
Our conventions are such that:
\be
\widehat{\mathcal{O}}(s) = e^{ i s H} \widehat{\mathcal{O}}(0) e^{ - is H}
\ee
The $i \epsilon$ makes the sum over intermediate energy eigenstates convergent. Note that
while the answer \eqref{cont} does not look real it actually is due to the integral over $\tau_a$ and
$\tau_b$ and symmetry under  $a \leftrightarrow b$.

We now want to calculate the EE.
Acting with $(1- \partial_n)$ and taking $n=1$ we have:
\be
\label{commute}
\delta S_{EE} = \left. (1 - \partial_n) \ln Z_n \right|_{n=1}  = 
\int d\mu \int_{-\infty}^{\infty} \frac{ds}{2\pi}  \frac{1}{ \left(e^{s - i \tau_{ba}} -1\right)} 
\left. \partial_n A_n (s) \right|_{n=1} 
\ee
where:
\begin{align}
\label{hoo}
\frac{1}{2\pi} \left. \partial_n A_n (s) \right|_{n=1} & = -i {\rm Tr} \left[ e^{ - 2\pi  H} H \widehat{\mathcal{O}}_a(0) \widehat{\mathcal{O}}_b(s + i \epsilon)   \right]+i {\rm Tr} \left[ e^{ - 2\pi  H}H \widehat{\mathcal{O}}_b(s - i \epsilon)   \widehat{\mathcal{O}}_a(0)   \right]
\end{align}
We would like to undo the steps we followed
above for the Renyi entropy and manipulate this  expression to remove 
the integral over $s$ and write the answer simply in terms of a Euclidean thermal greens function evaluated only at imaginary times. As we will see now the ordering of $H$ in \eqref{hoo} complicates this and we will not be able to fully remove the $s$ integral. 

The next few steps will be done assuming $0 < \tau_{ba} < 2 \pi $.  We will then
use a different but related set of steps for $ -2 \pi < \tau_{ba} < 0$, to be explained below. This
will give us results which have nice time-ordering properties. Firstly we commute
$H$ through $\widehat{\mathcal{O}}_b(s - i \epsilon)$ in the second term of \eqref{hoo} via
\be
[ H, \widehat{\mathcal{O}}(s)] = - i \frac{d}{ds} \widehat{\mathcal{O}}(s)
\ee
We then use the KMS condition to shift  $ s \rightarrow s + 2 \pi i $ in this same term with the operator ordering reversed:
\begin{align}
\nonumber
\frac{1}{2\pi} \left. \partial_n A_n (s) \right|_{n=1}  = -i {\rm Tr} & \left[ e^{ - 2\pi  H} H \widehat{\mathcal{O}}_a(0) \widehat{\mathcal{O}}_b(s  + i \epsilon)   \right]+i {\rm Tr} \left[ e^{ - 2\pi  H}H   \widehat{\mathcal{O}}_a(0)  \widehat{\mathcal{O}}_b(s - i \epsilon + i 2\pi)    \right] \\
&  + \frac{d}{ds} {\rm Tr}\left[ e^{ - 2\pi  H}  \widehat{\mathcal{O}}_b(s - i \epsilon)  \widehat{\mathcal{O}}_a(0) \right]
\label{dA}
\end{align}
The integral
over $s$ of both terms in the first line of \eqref{dA} can now be written as a single contour integral
which we can then deform to pickup the pole at $s = i \tau_{ba}$ in \eqref{commute}. 
This partially achieves our goal of writing the answer in terms of a Euclidean thermal greens function, although
we are left with the last line of \eqref{dA} which cannot be further manipulated. 

For $-2 \pi < \tau_{ba} < 0$ we follow very similar steps although instead we manipulate the first
term in \eqref{hoo} by commuting $H$  through $\widehat{\mathcal{O}}_b(s + i \epsilon)$ and shifting $s \rightarrow s - 2 \pi i$. We can
then again use contour integration to pickup the pole at $s = i \tau_{ba}$ (now below the real axis.)
Altogether we get two contributions to EE, one where we have removed the $s$ integral:
\be
\label{htt}
\delta S_{EE}^{(1)} = 2\pi \int d\mu {\rm Tr} \left[ e^{ -2\pi H} H \mathcal{T} \left( 
 \widehat{\mathcal{O}}_a( i \tau_a ) \widehat{\mathcal{O}}_b( i \tau_b) \right) \right]
\ee
and the other from the commutator terms left over in the second line of \eqref{dA}:
\be
\label{tord}
\delta S_{EE}^{(2)} = \int d\mu \int_{-\infty}^{\infty} d s \frac{1}{ 4 \sinh^2(s-i \tau_{ba})/2)}
 {\rm Tr} \left[ e^{ -2\pi H}  \mathcal{T} \left( 
 \widehat{\mathcal{O}}_a( 0 ) \widehat{\mathcal{O}}_b(s - i \epsilon \, {\rm sgn} (\tau_{ba}) ) \right) \right]
\ee
where we have done an additional integration by parts on the $s$ integral. The time ordering in \eqref{tord} just fixes the correct operator ordering such that the $i \epsilon$ 
makes the sum over intermediate states convergent. 
Note that $\delta S_{EE}^{(2)}$ is real - as can be shown by relabeling the $\tau_a$ and $\tau_b$  integrals in $d \mu$ such that $\tau_a \leftrightarrow \tau_b$.

The two contributions to the EE we have identified will turn out to have a natural and distinct
interpretation in terms of holography. The first contribution $\delta S_{EE}^{(1)}$ is
rather easy to deal with and has appeared previously in similar perturbative calculations
of EE. However the second term has not appeared before and in a sense will be the
most interesting term. We start by considering $\delta S_{EE}^{(1)}$ where we can
further manipulate the integrand by making a conformal transformation to write it in terms of a CFT 3 point function on flat space:
 \be
 \label{before}
\delta S_{EE}^{(1)} =  \frac{1}{2} \g^2 \int_a \int_b 2\pi  \left<  H \mathcal{O}(a) \mathcal{O}(b) \right>_1  
\ee
Note that the time ordering is built into Euclidean CFT correlation functions however it is important
that we evaluate the integral defining $H$ in terms of the stress tensor over
the region $A$:
\be
H = \int_A d \Sigma^\mu \xi^\mu  T_{\mu \nu}
\ee
Other homologous regions which end on the boundary of $A$ would give different
answers, despite this being a conserved charge (in the CFT), because of the operator insertions. 
This prescription can be gleaned from the operator ordering in \eqref{htt}.
The form \eqref{before} has appeared previously in perturbative calculations of EE  \cite{Rosenhaus:2014woa}.
We will interpret this simply as the expectation value of the CFT modular hamiltonian
in the deformed theory at second order in perturbation theory which we write as:
\be
 \label{s1h}
\delta S_{EE}^{(1)} = \left. 2\pi \vphantom{\sum} \left< H \right>_\g \right|_{\mathcal{O}(\g^2)}
\ee
From now on we will drop the cumbersome notation $|_{\mathcal{O}(\g^2)}$ and it should
be understood our expressions are only valid at second order in perturbation theory.
It turns out that this term is divergent. The appropriate divergences can
be seen by writing out the three point function of the stress tensor and two operators
as appears in \eqref{before}. We go through this carefully in Appendix~\ref{app:divs1}. For now we note
that we can avoid IR divergences by picking:
\be
d/2 < \Delta < d  
\ee
although this forces on us a UV divergences which appears when all three operators
come close together (there is no divergence simply when $\mathcal{O}(a)$ comes close
to $\mathcal{O}(b)$ due to the appearance of the stress tensor in this correlator.) To cut this divergence off we simply deform the $a,b$ integration regions so that it never coincides with the
stress tensor. For example we can achieve this via:
\be
\label{regcut}
|x^0_{a,b}| > \delta 
\ee
and such a regularization gives rise to the divergence:
\be
\label{ddiv}
\delta S_{EE}^{(1)} = c' ( 2 \delta)^{d- 2 \Delta} \int_{A} d^{d-1}\vec{x} 2\pi \xi^0 \g^2 + \ldots
\ee
This term then scales as $R^{d}$ which would lead to a super-area law divergence for the EE. This
is clearly not physical, so it is fortunate that we will find in the next section a canceling divergence in
$\delta S_{EE}^{(2)}$. In order to get these divergences to cancel it turns out that
we need to make the exact same regularization cut \eqref{regcut} for the integrals in 
\eqref{tord}.

Let us now turn to manipulating $\delta S_{EE}^{(2)}$ given in \eqref{tord}. Clearly this term will just be fixed by the CFT 2 point function on flat space. 
For example the Euclidean greens function on $\mathbb{H}_{d-1} \times \mathbb{S}$ at $\beta = 2\pi$
is simply,
\be
\label{defc}
G_1( \tau; Y_a, Y_b ) = \frac{ c_\Delta}{ \left( - 2 Y_a \cdot Y_b - 2 \cos(\tau) \right)^{\Delta}}
\qquad c_\Delta = \frac{ 2 (\Delta -h)  \Gamma(\Delta) }{ \pi^h \Gamma(\Delta - h)}
\ee
We have picked a specific normalization for this two point function (and thus for our coupling $\g$)
which is inspired by AdS/CFT \cite{Klebanov:1999tb} and we have set $h = d/2$.
To find the function appearing in \eqref{tord} we need simply to analytically continue this to $\tau \rightarrow - i s + \epsilon$. Combining everything the final answer is the following
set of horrendous integrals:
\begin{align}
\nonumber
\delta S^{(2)}_{EE} &=   \frac{1}{2} \g^2 c_\Delta R^{2(d- \Delta)}  \int_{ \delta } d \tau_a d\tau_b dY_a d Y_b  \int_{-\infty}^{\infty} ds  \left. \vphantom{\frac{Y^I}{ Y^\Delta}} \right. 
\\ & \qquad \qquad  \qquad \quad \left(   \frac{( Y^I_a+ \cos \tau_a )^{\Delta - d} ( Y_b^I + \cos \tau_b )^{\Delta - d}}{4 \sinh^2\left((s  - i   \tau_{ba})/2 \right) ( - 2 Y_a\cdot Y_b
- 2 \cosh (s - i \epsilon \, {\rm sgn} (\tau_{ba})) )^{\Delta} } \right)
\label{feel}
\end{align}
where the integral over the points $a,b$ are cutoff via the constraint \eqref{regcut}.

\begin{figure}[h!]
\centering
\includegraphics[scale=.9]{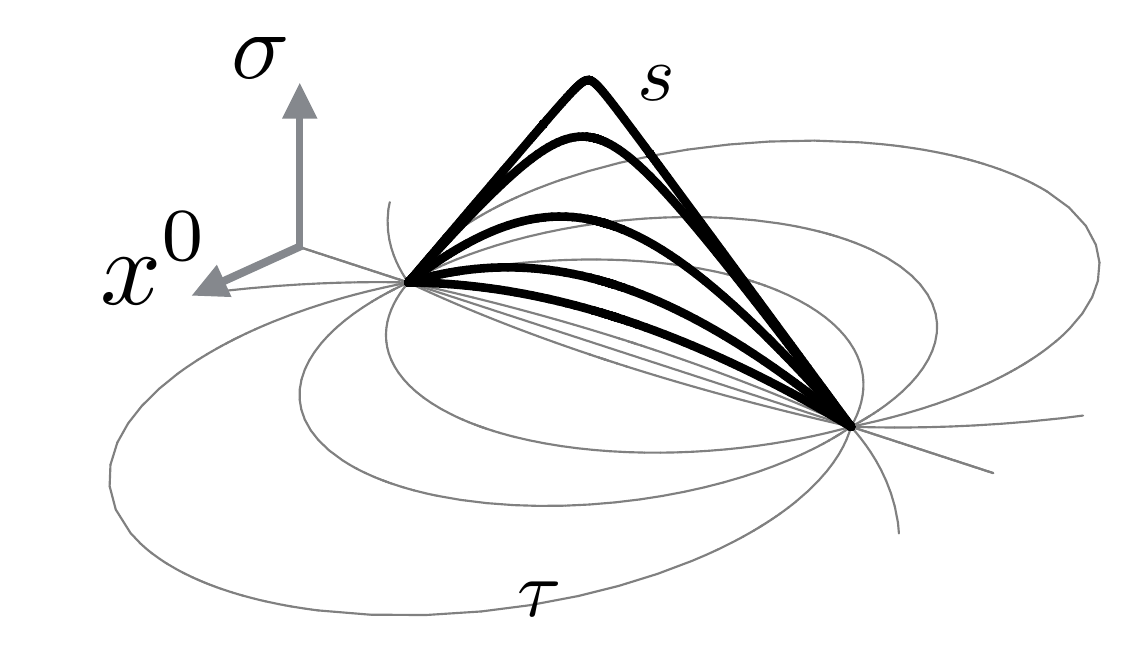} 
\caption{ \label{real-times} The calculation of EE forces us into real times, where we integrate
over the position of one of the operators after a real time modular flow transformation by an amount $s$. This modular flow pushes the operator into the causal development of $A$ pictured here (thick
lines are lines of constant $s$.) 
This is contrasted with the modular flow by an amount $\tau$ in Euclidean time (lighter lines are lines of constant $\tau$.) }
\end{figure}

To get some feeling for what \eqref{feel} means we manipulate a little further and write it in terms of
an integration over projective coordinates $P_a, P_b$. In order to do this it is convenience to introduce a spurious projective coordinate $P_\infty = \frac{1}{2R}(1,-1,0, \, \ldots  )$ which represents the point at infinity
for the flat coordinates on $P$. This
allows us to write expressions which respect conformal symmetry covariantly, although still being
broken by this fixed choice for $P_\infty$. We also deform the $s$ integration contour so it lies just above (below) the double pole at $s = i \tau_{ba} \mp 2 \pi i$ for $\tau_{ba} >0 ( \tau_{ba} <0)$ . This moves the $\tau_{ba}$ dependence
from the $\sinh^2()$ function to the two point function. We can then interpret the resulting term as a two point function between $P_a$ and $ M(i s) \cdot P_b$; the image of the second coordinate
under a modular flow in \emph{real} times (defined around \eqref{modflow}). 
 Together we find:
\be
\label{finint}
\delta S^{(2)}_{EE}  = \frac{1}{2} \g^2 c_\Delta  \int_\delta d P_a dP_b  \int_{-\infty}^{\infty} ds
\frac{ ( - 2 P_\infty \cdot P_a )^{\Delta-d} ( - 2 P_\infty \cdot P_b )^{\Delta-d} }{ 4 \sinh^2( (s + i \epsilon \, {\rm sgn} \tau_{ba} )/2) ( -2 P_a \cdot M( i s) \cdot P_b)^\Delta }
\ee 
Notice that the this forces us into real times where $M(is) \cdot P_b$ is some
point within the causal development of $A$ as shown in Figure \ref{real-times} (really this picture is only true for $\tau_b =0$, although
we think it is still a good picture to have in mind.) This is a highly non-local operation, especially
since we are integrating over the flow parameter $s$ with some kernel. 
We have found it very difficult to directly integrate this expression, many attempts by the author led to a dead end. Finally we found one method that works which we present next.

\section{The Integrals} 
\label{integrals}

We now set out to do the integrals in \eqref{feel}. We will use many tricks that may seem very ad-hoc. The reason
for this is that we worked some of the steps out in reverse, working backwards from an answer which was obtained using holography. We will present the calculation in the other way because we hope it highlights how holographic aspects emerge from a purely field theoretical calculation. We also
note in passing that these manipulations were indirectly inspired by some of the calculations in \cite{Penedones:2010ue}.

 For now we will only manipulate the terms depending on $s$ in \eqref{feel}.
 That is consider the $s$ integral:
 \be
 \label{inI}
 I \equiv  \int_{-\infty}^{\infty} ds \frac{1}{4 \sinh^2( ( s - i \tau_{ba})/2)( - 2 Y_a \cdot Y_b - 2 \cosh (s  - i \epsilon \, {\rm sgn} (\tau_{ba})) )^\Delta} 
 \ee
Let us enumerate some properties of $I$ that will be important for us later. 
Firstly note that complex conjugation is equivalent to sending $\tau_a \leftrightarrow \tau_b$. 
Secondly for fixed $Y_a \neq Y_b$ the function $I(\tau_a,\tau_b)$ is analytic in the region   $ 0 < \tau_a < 2 \pi$ and $ 0 < \tau_b < 2\pi$.  This is true except at the boundaries of this region where, under the periodic identification of $\tau_a$ and $\tau_b$, $I(\tau_a,\tau_b)$ suffers from cut discontinuities. For example one can easily check that the function is analytic
around $\tau_{ba} = 0$ via a simple contour manipulation of the $s$ integral, see Figure~\ref{contdef}
for an explanation of this.

\begin{figure}[h!]
\centering
\includegraphics[scale=.75]{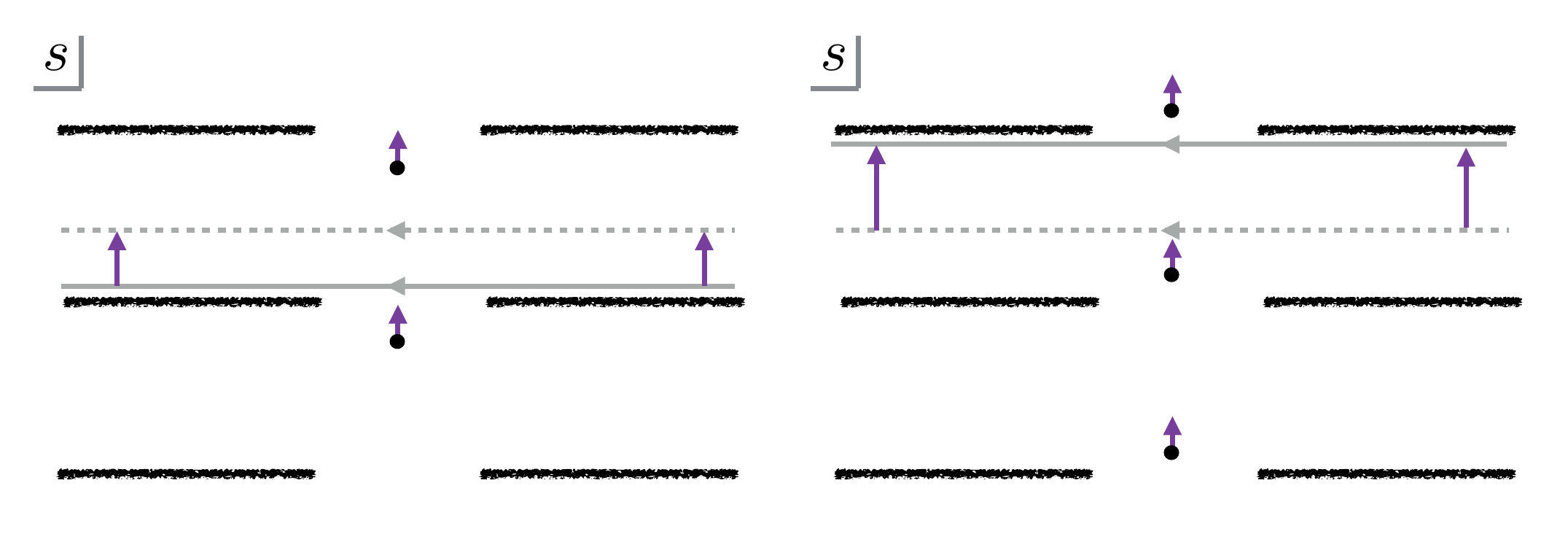} 
\caption{ \label{contdef} Pictorial argument showing that $I(\tau_a,\tau_b)$ is analytic near $\tau_{ba} = 
\tau_b - \tau_a = 0 $. The position of the poles in these pictures lies along the imaginary
axis displaced from the vertical location of the branch cuts by an amount $\tau_{ba}$. As we move
 $\tau_{ba}$ towards zero from below, we come close to the prescirbed integration contour in \eqref{inI} for $\tau_{ba} < 0$ at  $s = i \epsilon$.  We can then
deform this contour upwards towards $s = 2\pi i - i \epsilon$ being careful not to hit 
the next pole above the integration contour.  Once we have done this we can use the imaginary
time periodicity of the integrand to arrive at the prescribed integration contour in \eqref{inI} for
$\tau_{ba}>0$. }
\end{figure}

We need this last property because we are about to make some manipulations where
we cannot track the full function $I$ in the desired $(\tau_a,\tau_b)$ region - we will start
working in a small neighborhood\footnote{ Not infinitesimally small, actually
$ \pi/2 < \tau_{a,b} < 3\pi/2$ will do.} around $\tau_a\approx \tau_b \approx \pi$ and then use the above analyticity property
to move us outside of this region. 

We start by deforming the integration contour in $I$ from $s \rightarrow s - i \pi {\rm sgn} (\tau_{ba})$. Note that, as discussed in the paragraph above, to begin with we only consider $\tau_{ba} \approx 0$ 
and so we don't have to worry about this contour deformation passing one of the double poles at $ s = i \tau_{ba} + i 2 \pi m$. 
This first step is rather natural because it makes the arguments of the term raised to the power $\Delta$ real and positive, thus removing any confusions about which branch to take. We then have
for both signs of $\tau_{ba}$:
\be
I =  - \int_{-\infty}^{\infty} ds \frac{1}{4 \cosh^2( ( s - i \tau_{ba})/2)( - 2 Y_a \cdot Y_b + 2 \cosh s )^\Delta} 
\ee
We redefine $ \beta \equiv e^s$ and additionally exponentiate the power function using a Schwinger parameter:
\be
I = -  \frac{e^{i \tau_{ab}}}{\Gamma(\Delta)} \int_0^\infty \frac{d \beta}{\beta}  \int_0^\infty \frac{dt}{t}  \frac{\beta t^{\Delta}}{(\beta  + e^{i \tau_{ba} })^2} \exp\left( -  t ( \beta +\beta^{-1} - 2 Y_a \cdot Y_b ) \right)
\ee
We now change integration variables from $(\beta,t)$ to $(t_a,t_b) = (\sqrt{t/\beta},\sqrt{t \beta})$, in this way we can write the argument of the exponential suggestively as:
\be
t( \beta + \beta^{-1} - 2 Y_a \cdot Y_b ) = - ( t_a Y_a + t_b Y_b)^2
\ee
where we have used $Y_{a,b}^2 = -1$. We have:
\be
\label{denom}
I =  - \frac{ 2 e^{- i \tau_a - i \tau_b }}{ \Gamma(\Delta)} \int_0^\infty \frac{d t_a}{t_a}
\int_0^\infty \frac{d t_b}{t_b} (t_a t_b)^{\Delta + 1} \frac{ \exp\left( ( t_a Y_a + t_b Y_b)^2 \right)}
{ ( t_a e^{ - i \tau_a}  + t_b e^{ - i \tau_b})^2 }
\ee
The next step actually introduces bulk coordinates, although this may not be clear at this
point we will emphasize this connection already by labeling these coordinates appropriately. We
firstly use another Schwinger parameter $\ell_B$ to exponentiate the denominator in the last term of \eqref{denom}:
\be
\label{rep1}
\frac{1}{\left( t_a e^{-i \tau_a} + t_b e^{ -i \tau_b} \right)^2 } =
\int_0^\infty d \ell_B \ell_B \exp\left( \ell_B (t_a e^{-i \tau_a} + t_b e^{ -i \tau_b})  \right)
\ee
This last integral only converges  when
${\rm Re}( t_a e^{-i \tau_a} + t_b e^{-i \tau_b}) < 0$, 
which is satisfied for $\tau_{a,b}$ close to $\pi$. Of course 
we can work in this region of convergence, as we do for now, and simply analytically continue outside of it as necessary. We also introduce another coordinate $Y_B$ on $\mathbb{H}_{d-1}$ via:
\be
\label{yb}
 \exp\left( ( t_a Y_a + t_b Y_b)^2 \right)
  \rightarrow d_\Delta \int_{\mathbb{H}_{d-1}} \hspace{-.2cm} d Y_B \exp\left( 2 Y_B \cdot ( t_a Y_a + t_b Y_b )\right)
\ee
where $d_\Delta = \pi c_\Delta/(2 \Gamma(\Delta)(\Delta-h)^2) $ for $c_\Delta$
defined in \eqref{defc}. 
The reason we write only an arrow is because this manipulation is only valid under the
$t_{a,b}$ integrals in \eqref{denom}.
The easiest way to demonstrate this is then to work backwards.
Integrating the RHS of \eqref{yb} by rotating to $W \equiv t_a Y_a + t_b Y_b = |W|(1,0, \ldots)$
and using Poincare coordinates on $\mathbb{H}_{d-1}$ for $Y_B$ this becomes:
\be
d_\Delta \int \frac{d z d^{d-2} \vec{x}}{z^{d-1}} \exp\left( - |W| \frac{1+z^2+ \vec{x}^2}{z} \right)
\ee
We then rescale $(z,\vec{x}) \rightarrow (z,\vec{x})/ |W|$. After further rescaling $t_{a,b} \rightarrow t_{a,b} \sqrt{z} $ (and $\ell_B \rightarrow  \ell_B/\sqrt{z}$)
such that $W \rightarrow W \sqrt{z}$ we arrive at:
\be
\rightarrow d_\Delta \exp( W^2 ) \int \frac{ dz d^{d-2} \vec{x}}{z^{d-1}} z^{\Delta} \exp( -(z^2+ \vec{x}^2 )/z )
\ee
This last integral can be done and thus fixes the constant $d_\Delta^{-1}$ which we gave above.

\begin{figure}[h!]
\centering
\includegraphics[scale=.9]{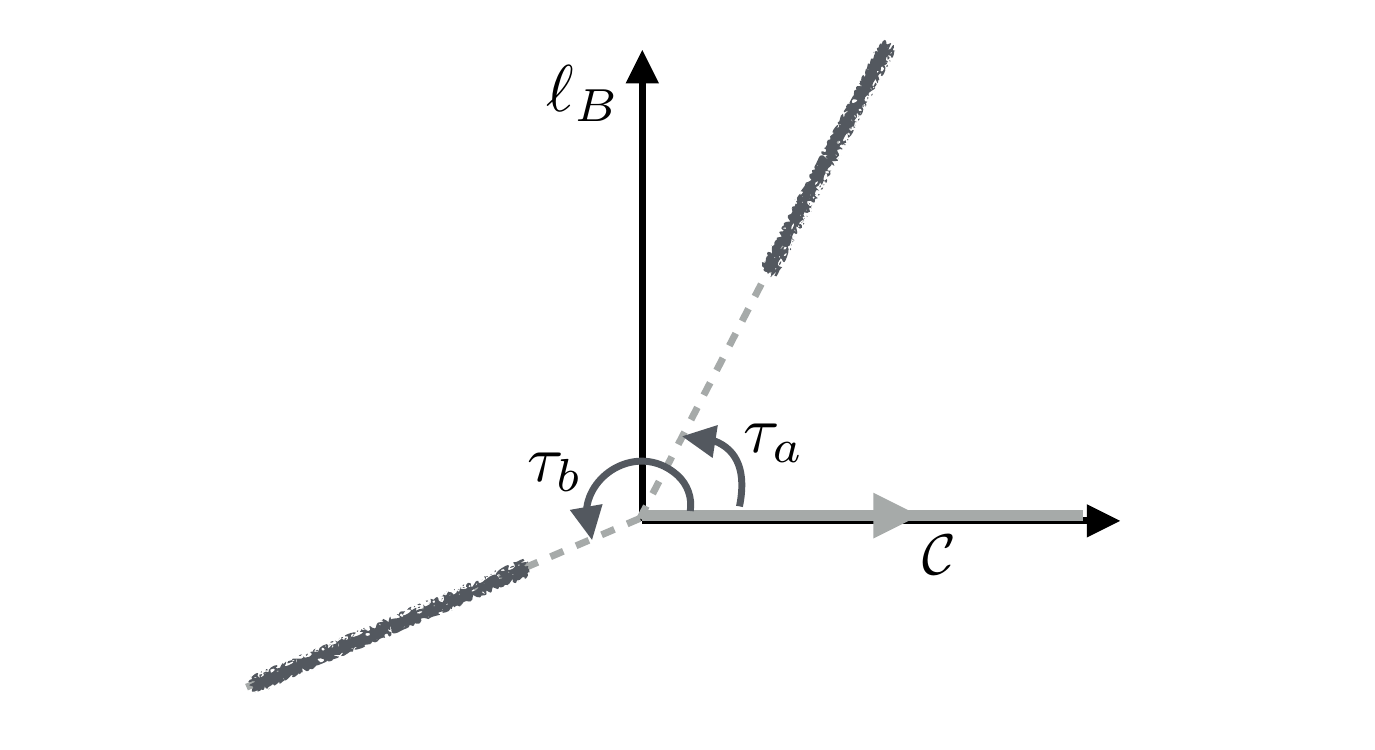} 
\caption{ \label{fig:ellb} The $\ell_B$ integral in \eqref{finI} in the complex $\ell_B$ plane 
at fixed $Y_B$. We denote
the integration contour by $\mathcal{C}$. 
 The two branch points are located at $e^{i \tau_{a,b}}/(-2 Y_{a,b}\cdot Y_B)$ . As long as these branch points do not cross the real axis there is no discontinuity and so the answer is analytic in the domain $ 0 < \tau_{a,b} < 2\pi$.
}
\end{figure}

After making the above two replacements (\ref{rep1}-\ref{yb}) and then integrating over $t_a$ and $t_b$ now that they appear linearly in all the exponential terms we find the tantalizing expression:
\be
\label{finI}
I =  - \frac{\pi  c_\Delta}{(\Delta - h)^2}  \int_0^\infty \hspace{-.3cm} d \ell_B \ell_B \int \hspace{-.1cm} d Y_B
\left[ 
\frac{\partial}{\partial \ell_B } \frac{1}{ \left( - 2 Y_a \cdot Y_B - \ell_B e^{- i \tau_a} \right)^{\Delta }} 
\right] \left[ 
\frac{\partial}{\partial \ell_B}\frac{1}{\left( - 2 Y_b \cdot Y_B - \ell_B e^{- i \tau_b} \right)^{\Delta}}
\right]
\ee
Before going into details about how to interpret $\ell_B, Y_B$ we quickly 
address the analytic properties of \eqref{finI} in the $\tau_a,\tau_b$ plane. Recall
that we made the above manipulations on $I$ assuming $\tau_a,\tau_b$ were both close to $\pi$.
However it is not hard to see that \eqref{finI} can easily be 
continued outside of this region  to $0 < \tau_{a,b} < 2 \pi$. This was the desired region of analyticity for the original
function in \eqref{inI} and so indeed \eqref{finI} should be our final expression for $I$. 
See Figure~\ref{fig:ellb} for pictures describing this. We note that these analyticity requirements 
fix the $\ell_B$ integration contour in the complex $\ell_B$ plane uniquely to lie on the positive real axis. Any other choice would necessarily give non-analyticities away from the boundaries at $\tau_a, \tau_b = 0, 2\pi$.

\begin{figure}[h!]
\centering
\includegraphics[scale=1]{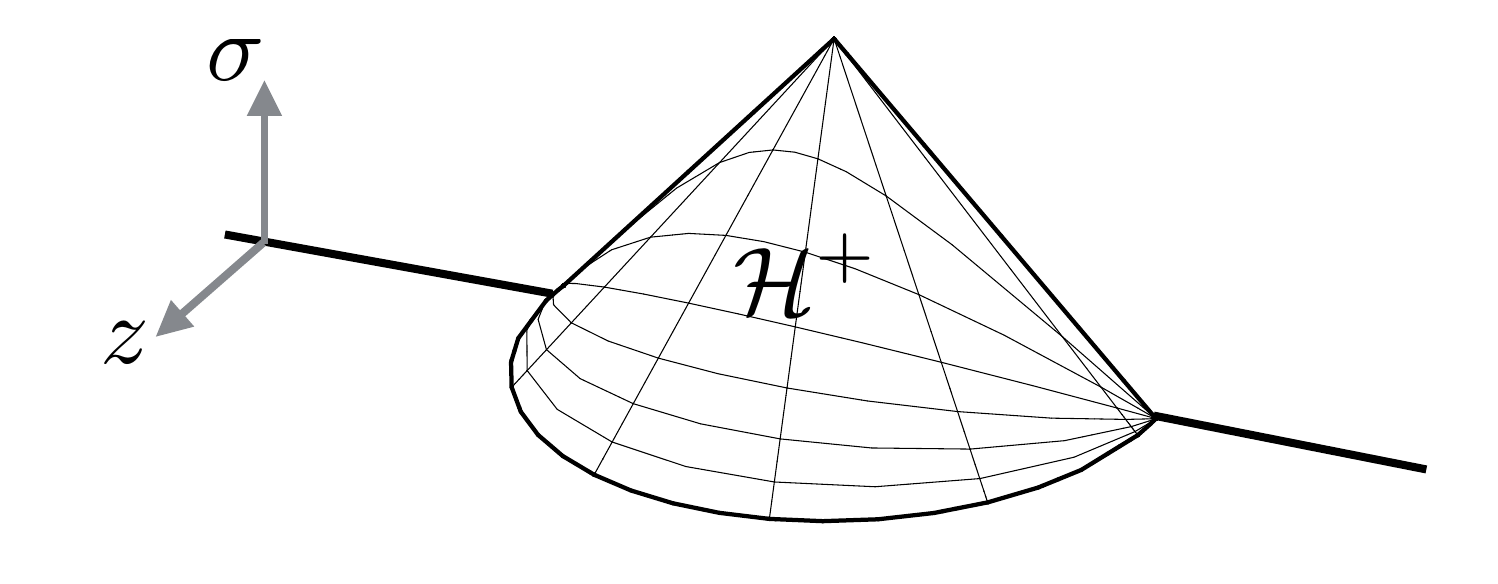} 
\vspace{-.3cm}
\caption{ \label{fig-hplus} Future part of the bulk rindler horizon. This is also the horizon
associated to the hyperbolic black hole which was discussed in \cite{Casini:2011kv}.}
\end{figure}

As we will show in the next few paragraphs 
the coordinates we introduced $\ell_B$ and $Y_B$  in \eqref{finI} parameterize the 
Rindler like horizon $\mathcal{H}^+$ in the bulk of an emergent $AdS_{d+1}$ space. This region 
can also be thought of as the horizon of the hyperbolic black hole introduced in \cite{Casini:2011kv}. So for example $\mathcal{H}^+$ ends  on the boundary of $AdS$ at the boundary of the future part of the  causal diamond associated to region $A$, see Figure~\ref{fig-hplus}. 
The $\ell_B$ coordinate
is an affine parameter along generators of $\mathcal{H}^+$.  These generators cover the future part  the horizon subtending from the bifurcation point at $\ell_B = 0$. This point lies on the minimal surface $m(A)$ associated to region $A$ in the bulk via the RT prescription \cite{Ryu:2006bv}.  
Then $I$ is simply an integral over $\mathcal{H}^+$ weighted by what will turn out to be the null energy of an emergent bulk field.

We go through  the identification of the bulk integral in more detail now. To do this
we have to introduce a few more ingredients.
Recall that the embedding space formalism allowed us to place coordinates
on various conformally flat $d$ dimensional spaces. 
This formalism is also useful for studying  $AdS_{d+1}$ itself, which for now
we take to be the Euclidean version, that is $d+1$ dimensional hyperbolic space.
This is defined via the hyperbola in embedding space:
\be
\label{hyp}
X^2 = - (X^I)^2 + (X^{II})^2 + X^\mu X^\mu = -1 \qquad X^I > 0
\ee
We then recover the projective coordinates $P$ by examining the conformal boundary of \eqref{hyp} which is where the conformally flat $d$ dimensional space lives. 
We can introduce coordinates on (Euclidean) AdS which then naturally limit to the various coordinates
we chose for $P$. The two cases of interest are respectively, Poincare coordinates and
hyperbolic black hole coordinates:
\begin{align}
X &= \left( \frac{R^2 + x^2 + z^2}{ 2 R z } ,\frac{R^2 - x^2 - z^2}{ 2 R z } , \frac{x^\mu}{z} \right) \\
X &= \left( r Y^I,  \sqrt{r^2-1} \cos \tau, \sqrt{r^2-1} \sin \tau, r Y^m \right)
\end{align}
where as usual $Y$ parameterize a $\mathbb{H}_{d-1}$. Note that we recover
the projective coordinates by rescaling and taking respectively 
$z \rightarrow 0$  or $r\rightarrow \infty$:
\be
P|_F = \lim_{z \rightarrow 0} X z  \qquad P|_H = \lim_{r \rightarrow \infty} X/r
\ee

The bulk point of interested to us is most easily described using the hyperbolic
black hole coordinates, however now in real times. 
So firstly we wick rotate by setting $\tau = - i s$.
We take the near horizon limit as $ r \rightarrow 1$ and scale $ s\rightarrow  \infty$
as we do this, such that $ \sqrt{r^2-1} e^{s}= \ell_B$ is held fixed. This corresponds to
the point:
\be
X_B = \left( Y^I_B, \ell_B/2,  -  i \ell_B/2, Y^{m}_B \right)
\ee
It can fairly easily be seen in Poincare coordinates (and in real times $\sigma = i x^0$) that this corresponds (with $\ell_B >0$) 
to the light cone region $ \vec{x}^2 + z^2 = ( R - \sigma)^2$ for $\sigma > 0$.

We can then simply write the answer for this contribution to EE following from \eqref{finI} as:
\begin{align}
\label{bulkmod}
\delta S^{(2)}_{EE} & =  - 2 \pi 
\int d \ell_B \ell_B \int d Y_B  \left( \partial_{\ell_B} \phi \right)^2 \\
{\rm where} \qquad 
 \phi(X_B) &\equiv  \g  \frac{c_\Delta}{2 (\Delta - h)} \int d\tau d Y \frac{ ( Y^I + \cos \tau)^{\Delta -d} }{ \left( - 2 Y \cdot
  Y_B - \ell_B e^{ - i \tau_A} \right)^\Delta}
  \label{tauint}
  \\ & = \g  \frac{c_\Delta}{2 (\Delta - h)} \int d P \frac{ (-2 P_{\infty} \cdot P )^{\Delta-d} }{ ( -2 X_B \cdot P )^{\Delta}}
 \label{dphi}
\end{align}
It should be clear that $\phi$ corresponds to an emergent dual field associated to the operator $\mathcal{O}$.
Note that the integrand in \eqref{dphi} is just the bulk to boundary propagator in $AdS_{d+1}$
for a scalar field of mass:
\be
m^2 = \Delta( \Delta-d)
\ee
So for example this means that:
\be
\label{kg}
\nabla^2 \phi - m^2 \phi  = 0 
\ee
where the covariant derivative is with respect to the background $AdS_{d+1}$ 
metric
(this is the naturally induced metric on the hyperboloid.)
Further it is not hard to recognize that the integrand in \eqref{bulkmod} is related
to the null component of the stress energy tensor of the bulk field $\phi$ 
integrated over the horizon $\mathcal{H}^+$.

We need to specify
the exact integration contour for the $P $ integral in \eqref{dphi}. 
The reason this is  little tricky follows from the fact that we were forced into real times
where one usually needs a prescription for dealing with such bulk to boundary propagators. This can be understood in the usual AdS/CFT language
as due to an ambiguity in the state of the theory in real times, which needs to be specified
via boundary conditions. As we will see now a natural and intriguing prescription will be
forced upon us.

\begin{figure}[h!]
\centering
\includegraphics[scale=.8]{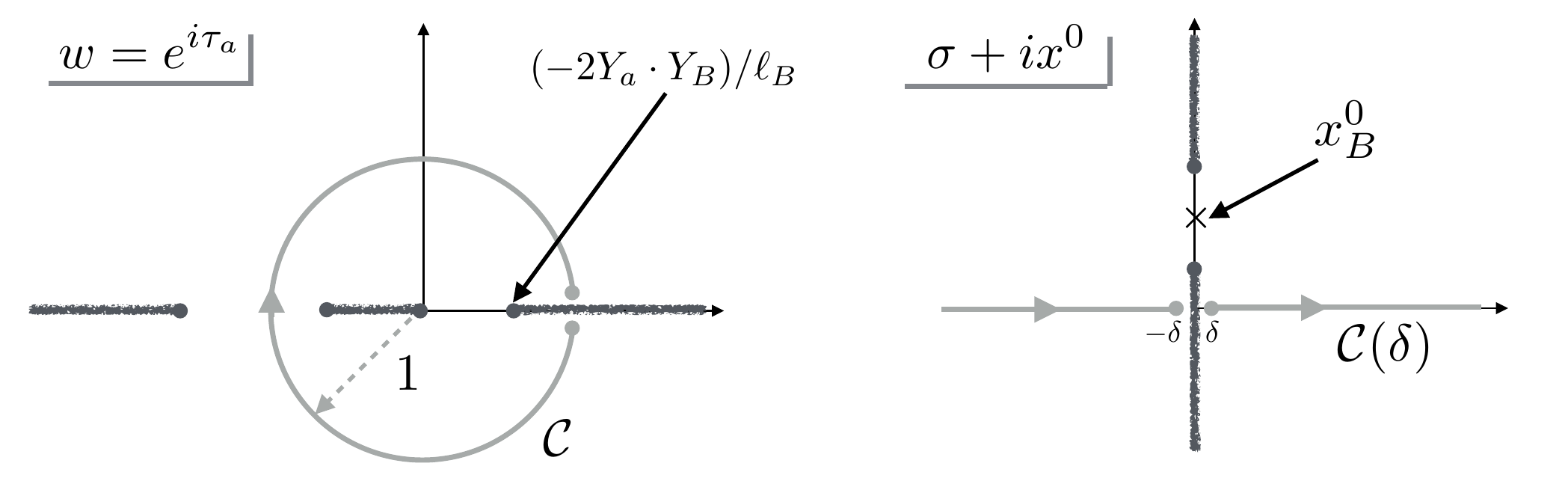} 
\caption{\label{fig:zplane} The real time contour prescription for calculating $\phi(X_B)$
either in hyperbolic coordinates (left) or poincare coordinates (right). Not only do we have
to make a jump  across different branches of the integrand, we should do this
slightly away from the branch cut as determined by the UV cutoff $\delta$. When evaluating
this integral for bulk coordinates in the euclidean section no such complications arise since
the contour never hits the branch cut.  }
\end{figure}

The integral over $\tau$ in \eqref{tauint} can be written as a contour integral in the complex
$w = e^{i \tau}$ plane circling the origin with radius $1$.  The structure of this
complex plane is shown in Figure~\ref{fig:zplane}. Note that for $\ell_B > -2 Y \cdot Y_B $ 
the $w$ integration contour is no longer closed due to a branch cut.
As we discussed already for $I$ in \eqref{inI} we expect a non-analyticity 
at $ \tau_{a,b} = 0 , 2\pi$ and so this fact fixes our choice for the branch cut in the $w$ plane to lie exactly along the real $w$ axis (any other choice would give a different result for this integral.) 
Further to this, recall that in order to tame an expected UV divergence we should cutoff the $a,b$
integrals to avoid coincidence of these points with the region $A$ which is located in hyperbolic coordinates at $\tau_{a,b} = 0, 2 \pi$  (see the discussion
around \eqref{regcut}.) So this divergence comes from integrating through the branch point
in Figure~\ref{fig:zplane} and the prescription in
\eqref{regcut} will regulate the divergence.\footnote{In $S^1 \times \mathbb{H}^{d-1}$ coordinates we should
cutoff the integral at $\tau_a > \delta_H $ and $\tau_a < 2\pi -\delta_H$
where  $\delta_H = \delta (Y^I_a + 1)/R$ in order to match
the flat space cutoff identified in \eqref{regcut}. Actually any reasonable cutoff choice should
work as long as we do this consistently for the integrals in $\delta S^{(2)}_{EE}$ and 
$\delta S^{(1)}_{EE}$. }

We now translate this prescription for the $P$ integral into the flat space coordinates. 
We also take $X_B$ to be in Poincare coordinates for $AdS_{d+1}$, and we should remember to wick
rotate $X_B$ to real times, as is appropriate for the points lying on $\mathcal{H}^+$:
\be
\label{bbprop}
\phi(X_B) = \g  \frac{c_\Delta}{2 (\Delta - h)} \int d^{d-1} \vec{x} \int_{\mathcal{C(\delta)}} dx^0 \frac{ z^{\Delta} }
{ \left( z^2 + (x^0 - i \sigma)^2 +  ( \vec{x} - \vec{x}_B)^2 \right)^\Delta} 
\ee
where we have defined the real time bulk coordinate $\sigma$ via $x^0_B = i \sigma$.
Again we have to be careful about our contour choice for the $x^0$ integral which 
we denote $\mathcal{C}(\delta)$.
Here we find that we must place the branch cut in the $x^0$ plane along the imaginary axis as in Figure~\ref{fig:zplane} and the $x^0$ integral should jump from one branch to another when
$x^0 = \pm \delta$.

The integrals defining $\phi$ can be done, and we give the answer
in Poincare coordinates:
\be
\label{phiint}
\phi(X_B) = \g z^{d-\Delta} F_\delta( \sigma/z) 
\ee
where the scaling function is defined as:
\begin{align}
\label{Ffirst}
&F_\delta( q) \equiv c' \int_{\mathcal{C}(\delta/z)} d y \left( 1+ (y - i q)^2 \right)^{-\Delta + h - \half} \\
\label{Fsecond}
& =  \left\{ \begin{matrix}  1, &  \quad 0 < q < 1 - \mathcal{O}(\delta/z) \\ 
 1- \frac{\sqrt{\pi} (q^2-1)^{-\Delta + h +\half}  } { q \Gamma(\Delta - h)}  \left._2 F \right._1 \left( \half,1; \frac{3}{2} + h -\Delta;
 1 - \frac{1}{q^{2}} \right), &  \quad  q > 1 + \mathcal{O}(\delta/z)
  \end{matrix} \right.
\end{align} where $c'$ is the same constant as that defined in \eqref{cpdef}.
In \eqref{Fsecond} the $\delta$ cutoff smooths out the singular behavior in the region between the 
limits quoted, the
precise form of $F_\delta$ in this region can be worked out by following the contour
prescription $\mathcal{C}(\delta/z)$ as defined in the right part of Figure~\ref{fig:zplane}.

Let us note a few features of this answer. Firstly as long as we pick $X_B$ to be in the imaginary
time section we always get the simple answer $\phi = \g z^{d-\Delta}$. Such an answer is to be expected
from the usual AdS/CFT dictionary where this corresponds to the leading falloff behavior for general
solutions to \eqref{kg}. This term is  interpreted as the coupling, which in
this case is a uniform deformation.
The reason the sub-leading (vev)  term ($\propto z^{\Delta}$) is not
present is due to  the fact that we are working in perturbation theory
where it is not possible to generate a non-zero vev for a uniform  deformation
(the vev one finds for example in the domain wall flows \cite{Bianchi:2001de} is always non-perturbative/non-analytic in $\g$.)

However away from imaginary times beyond the the region $\sigma > z$ we find
a correction to this answer. We interpret this as due to a quench - our prescription (which
was \emph{forced} on us) effectively sets the coupling $\g$ to zero in real times. This quench
then only effects the bulk beyond the region of causal influence $\sigma > z$. We could set up this
problem as an initial value problem, where $\phi(X_B)$ in Euclidean times sources the time
evolution in real times after imposing that $\g = 0$ at the boundary for this subsequent time evolution. 
The answer we would find is \eqref{phiint}. For example if we expand near the boundary
in real times we have:
\be
\label{vevtime}
\phi(X_B) =  \frac{2 \cos( \pi (\Delta -h)) \Gamma(2 \Delta - d ) }{ (\Delta -h)\Gamma (\Delta -h)^2 } \g | 2 \sigma|^{d - 2\Delta}  z^{\Delta} + \ldots
\ee
which describes the evolution of the vev in real times. Note the answer is divergent
as $\sigma \rightarrow 0$ and is cutoff at $\sigma \approx \delta$.
This divergence seems to be related to a result found in \cite{Buchel:2013gba} for fast quenches.

Note that the $\ell_B$ integral  for $\delta S^{(2)}_{EE}$  in \eqref{bulkmod} is actually
divergent when plugging in the solution \eqref{phiint} - and this divergence is regulated by the $\delta$ cutoff. This can be seen by making the substitutions back to hyperbolic coordinates
$\sigma/z = \ell_B/2$ and $ R/z = (Y^I_B+\ell_B/2)$. The divergence is then seen
to occur around $\ell_B \sim 2$ for $\Delta > d/2$. Naively it seems that this divergence is not a UV
divergence, since it occurs well away from the boundary, however we will now show that
we can push the divergence all the way up to the boundary where we will be able to identify
it as the canceling divergence that kills the problematic term \eqref{ddiv}.

To proceed we would like to avoid the difficulties associated with working in real times, and the
quench prescription that comes along with this. In order to do this we start by
pushing the bulk integral over $\mathcal{H}^+$
away from the horizon down to a region where $\sigma = 0$. We are allowed to do this because the integral in \eqref{bulkmod} defines a conserved charge in the bulk, it can be written generally as:
\be
\label{newexp}
\delta S_{EE}^{(2)} =
 - 2 \pi \int_{\mathcal{R}} d \Sigma^a \xi^b_B T_{a b}^B  
\ee
where we have introduced:
\be
T^B_{ab} = \partial_a \phi \partial_b \phi - \frac{1}{2} \left( m^2 \phi^2 + (\partial \phi)^2 \right) g_{ab}^0
\ee
where $g^0$ is the metric on $AdS_{d+1}$
and the bulk region $\mathcal{R}$ is some region homologous to $\mathcal{H}^+$. Any two such regions gives the same answer because we constructed
$T_{ab}$ to be conserved $\nabla^a T_{ab } = 0$ and because $\xi_B$
is a bulk killing vector for the $AdS$ metric. That is:
\be
\xi_B =\frac{R^2 - \vec{x}^2 - \sigma^2 - z^2}{2R}  \partial_\sigma   - \frac{\sigma}{R} ( \vec{x} \cdot \vec{\partial} + z \partial_z )
\ee
This is the killing vector that limits to the conformal killing vector $\xi$ on the boundary.
We take coordinate labels $a,b$ to be in real times and have defined $\xi_B$ in real times.

\begin{figure}[h!]
\centering
\includegraphics[scale=.9]{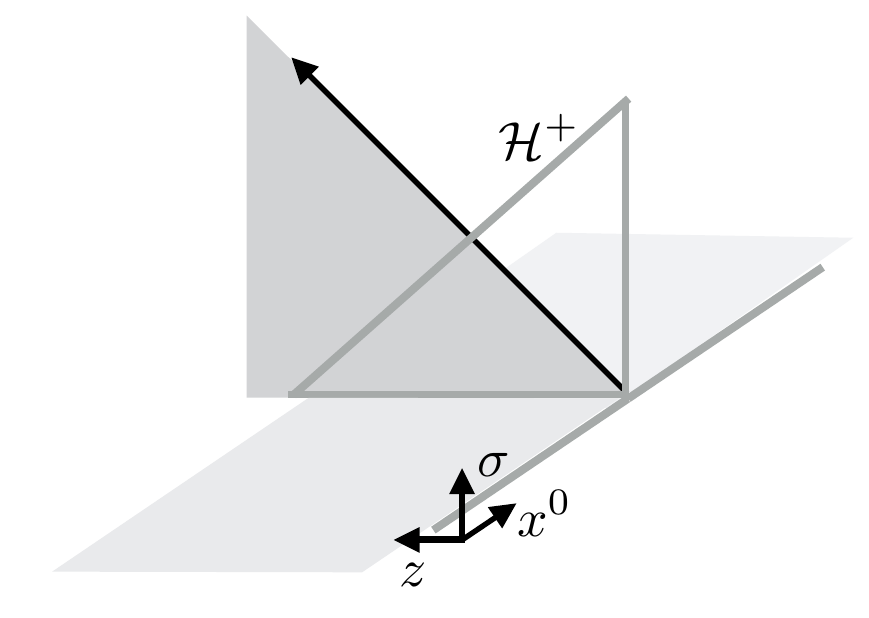} \hspace{.5cm} \includegraphics[scale=.9]{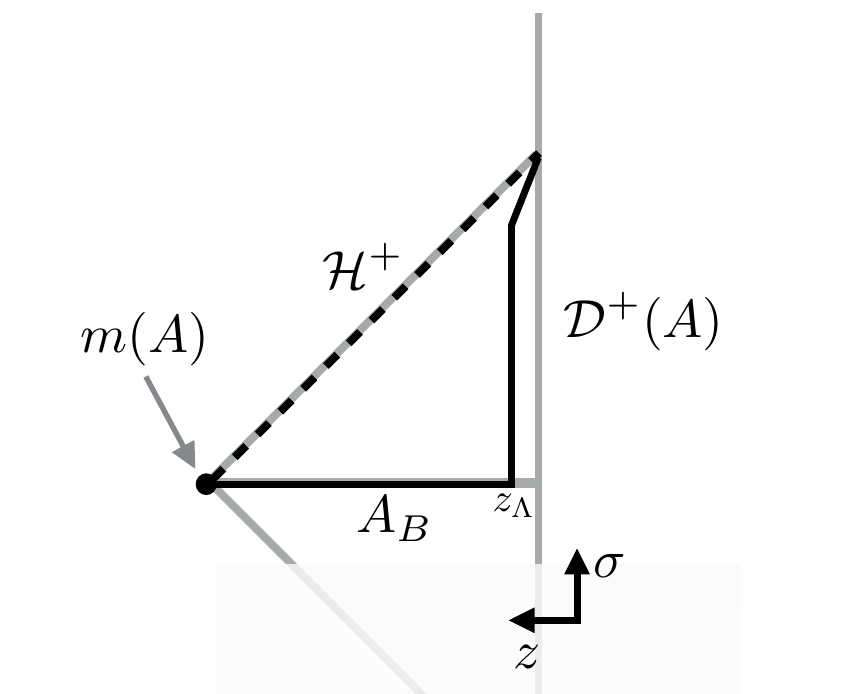} 
\caption{\label{fig:tworegions} (\emph{left}) A picture of the quench prescription
derived from the EE calculation, showing both imaginary times and real times matching
along the surface $\sigma = x^0 = 0$.  The arrow represents a shockwave due to the quench
in the coupling $\g$ which is set to zero moving into real times. The bulk solution for $\phi$ remains time independent in the shaded region.  (\emph{right})  We deform the bulk integral over $\mathcal{H}^+$ away
from the null surface and get two contributions, one from the spatial integral over $A_B$ and the other from close to the boundary at $z=z_\Lambda$ in $\mathcal{D}^+(A)$ the future domain of dependence
of $A$ on the boundary. }
\end{figure}

It can easily be checked that \eqref{newexp} gives the same answer as \eqref{bulkmod} by
pushing the region $\mathcal{R}$ to the lightcone  $\mathcal{H}^+$, where we get
the required integral of the null stress energy of $T^B$ over the horizon generators. We can now push $\mathcal{R}$ down to $\sigma = 0$, although we must additionally take care of a ``vertical'' contribution from close to the boundary at $z = z_\Lambda$ (see Figure~\ref{fig:tworegions}):
\be
\label{regent}
\delta S_{EE}^{(2)} =   - 2 \pi \int_{A_B} 
d \Sigma^a \xi_B^{b}  T_{ab}^B
  +
  2\pi z_\Lambda^{1-d} \left. \int_{\mathcal{D}^+(A)} d^{d} x   \xi^a_B  T^B_{a z}\right|_{z = z_\Lambda} 
\ee
where we have defined the bulk region $A_B$ to be a region on the space-like surface $\sigma = 0$ that lies between the RT surface $m(A)$ and the region $A$ cutoff close to the boundary at $z=z_\Lambda \rightarrow 0$. 
We need to do this $z_\Lambda$ regularization because the two individual terms above are separately
divergent as we take $z_\Lambda \rightarrow 0$ but these divergences cancel between the two terms. 

The second term in \eqref{regent} involves
an integral over the boundary in the future causal development of $A$
as shown in more detail in Figure~\ref{fig:tworegions}. 
We evaluate this term in detail in Appendix~\ref{app:ct}. 
 For $\sigma$ fixed as $z_\Lambda \rightarrow 0$
there is no contribution since the leading behavior $\phi \sim z^\Delta$ given in \eqref{vevtime}
can never contribute a finite or divergent term as $z_\Lambda \rightarrow 0$
(for $\Delta > d/2$.)
Thus the main contribution comes from $0 < \sigma  \lesssim z_\Lambda$ including the promised
$\delta$ divergent term (we take $\delta \ll z_\Lambda$):
\be
\label{ct1}
S_{ct} = 2\pi \int_A d^{d-1} \vec{x} \xi^\sigma \left(  - \frac{( d- \Delta)}{2} \g^2 z_\Lambda^{d-2 \Delta}
- c' \g^2 (2 \delta)^{ d- 2 \Delta} \right) 
\ee
We will refer to $S_{ct}$ as an ``entropy counter term''. Note that the second
term above exactly cancels the expected divergence in $\delta S_{EE}^{(1)}$.

We could continue interpreting this result in terms of holography, however we choose now to continue
to blindly evaluate the integral setup in \eqref{regent}. We will return to holography in the next section.
\begin{align}
\delta S^{(2)}_{EE}  & = 
-  \pi \int_{A_B} dz d^{d-1} \vec{x} \sqrt{-g} \xi_B^\sigma \nabla_\mu ( \phi \nabla^\mu \phi)   + S_{ct}
\\
&=  \pi( d-\Delta)(2\Delta - d) \g^2 S_{d-2} \int_{z_\Lambda}^R
\frac{d z}{z}  z^{d- 2 \Delta}\int_0^{\sqrt{R^2-z^2}}  \hspace{-.5cm} d r r^{d-2} \frac{ \left( R^2 - r^2 - z^2
\right) }{2R}  + S_{ct}
\\
& = - \g^2 R^{2(d-\Delta)} \frac{ \pi^{h+\half} (d-\Delta) \Gamma( 1 + h - \Delta)}{ 2 \Gamma( \frac{3}{2} + d - \Delta)}  + \g^2 R^{d} z_\Lambda^{d - 2 \Delta} \frac{  \pi^{ h + \half}(d- \Delta)}{ 2\Gamma( \frac{3}{2} + h)} + S_{ct}
 \label{finalint}
\end{align}
The second term is divergent as $z_\Lambda \rightarrow 0$ and this term cancels one of the
terms in $S_{ct}$. Finally we note that apart from the $\delta$ divergence that we identified
for $\delta S_{EE}^{(1)}$ there are no other possible terms we can write down
for $\delta S_{EE}^{(1)}$. The argument
follows from scale invariance and the lack of any other scale in the problem. That is if we first calculate
$\left< T_{tt} \right>_\g$ at order $\g^2$ we can only find the divergent answer
$ \propto \g^2 \delta^{ d- 2 \Delta}$ with no other term possible in perturbation theory.  
After integrating over $A$ multiplying in $\xi^\sigma$ this divergent term cancels the remaining term in $S_{ct}$ and in totality we are left with a finite answer, that is the first term in \eqref{finalint}:
\be
\delta S_{EE} = - \g^2 R^{2(d-\Delta)} \frac{ \pi^{h+\half} (d-\Delta) \Gamma( 1 + h - \Delta)}{ 2 \Gamma( \frac{3}{2} + d - \Delta)}
\ee
For example if we set $d=3$ and calculate the $F$ function as in \eqref{Ffunc} we
reproduce the term \eqref{checkhong}. For other dimensions we make similar predictions
which all agree with the holographic results in \cite{Liu:2012eea} and also \cite{Nishioka:2014kpa}.\footnote{Note that in order
to make the comparison exactly some extra work is needed to translate to the conventions
of these papers. In particular the scale $\mu$ that appears in \cite{Liu:2012eea}  can be
shown to be related to our coupling via $ \mu^{2(d-\Delta)}/8\pi G_N = \lambda^2 (d-\Delta)/(d-1)$. } Thus we have extended these results beyond holography.

At this point we would like to emphasize that we have managed to do the integrals \eqref{feel} that we set out to do.
More direct attempts by the author at this integral have not been successful, although that does not mean there is not a more direct method. We can thus see the appearance of holography and the dramatic simplifications that occur from this perspective as a potentially very useful tool in studying EE in any QFT. That is independent of the tantalizing hints of bulk emergence, the methods introduced here should be useful for a wide variety of problems in the study of EE.

\section{Bulk Emergence}
\label{emergence}

Firstly we note an immediate generalization, which is to spatially dependent couplings.
That is we would like to deform the theory by:
 \be
 S_{CFT} \rightarrow S_{CFT} + \int d^d x \g(x) \mathcal{O}(x)
 \ee
 For now we stick to deformations of the  Euclidean theory. There should be no obstruction to working in real time with dynamics, especially in perturbation theory, but we leave this to future work.   
 One issue with the restriction to Euclidean is that we may not actually be able to interpret our results
 as a calculation of EE, in the sense that there is some hermitian reduced density
 matrix acting on some Hilbert space. A condition that should be imposed in order
 to get an EE interpretation is the existence of
 a time reflection symmetry about the Entangling surface. For general $\g(x)$ this may not
 be true.  Instead we \emph{define} EE in the Euclidean theory
 simply via a correlation function of a twist surface operator. 
This is similar to the generalized notion of entropy discussed in \cite{Lewkowycz:2013nqa} in the context of theories with classical gravity descriptions.
 
Despite the emphasis on the Euclidean theory, 
in this section we will use a hybrid notation for our coordinates where we take the bulk coordinates to be  labelled by $(z,\sigma = i x^0, \vec{x})$ - effectively placing us in real times. 
Since we always work about the surface $\sigma =0$ this is achieved by a simple wick rotation with no extra complications - i.e. this does not \emph{really} move us into real times. 
We use these coordinates to allow for efficient comparison to previous work (and to keep track of signs via positivity requirements.) We can consider EE of rotated and translated regions by simply imagining rotating and translating the non-uniformity of the coupling $\g$, such that we always work with coordinates where the region $A$ lies on the surface $x^0 = \sigma=0$ centered on the point $\vec{x}=0$.
 
Following similar steps to the above section \footnote{
Basically the same calculation. The replica trick is achieved by putting the same coupling function $\g(x)$ on each replica and so for example for most of the calculation we can simply replace $\int d\mu $ with:
 \be
\int d\mu \ldots  \rightarrow  \frac{1}{2} \int_0^{2\pi } \hspace{-.2cm} d\tau_a  \int_0^{2\pi }  \hspace{-.2cm} d\tau_b 
\int_{\mathbb{H}_{d-1}} \hspace{-.5cm} dY_a  \int_{\mathbb{H}_{d-1}} \hspace{-.5cm}  dY_b  \, \, \Omega^{d-\Delta}(a)  \Omega^{d-\Delta}(b) \g(a) \g(b)
\ldots 
 \ee
 } we arrive at:
 \be
 \label{ee1}
 \delta S_{EE} = -  2 \pi \int_{A_B} d \Sigma^a \xi_B^b T_{ab}^B
 + S_{ct} + 2 \pi \left< H \right>_{\lambda}
 \ee
where the difference now is that we have
a new form for the bulk field $\phi$ given by:
\be
\phi(X_B) = \int_{C(\delta)} d^{d} x \g( x) K_\Delta(x, X_B)
\ee
where $K_\Delta$ is the same bulk to boundary propagator that appeared in 
\eqref{bbprop}. The contour prescription is the same as before, and a $\delta$ cutoff
is necessary to move $X_B$ into real times. Indeed we are still forced into real times but we used
the same tricks as in the previous section to move the integral over $\mathcal{H}^+$ to
the Euclidean section. This then generates the $S_{ct}$ terms which has a slightly different form
to the uniform coupling case:
\be
\label{ct2}
S_{ct} + 2 \pi \left< H \right>_\g = 2\pi \int d^{d-1} \vec{x}_B \xi^\sigma \left( - \Delta \beta \g - \frac{( d- \Delta)}{2} g^2 z_\Lambda^{d-2 \Delta}
+ \left< T_{\sigma\sigma} \right>^{\rm ren}_\g \right) 
\ee
where above $\g$ is evaluated at $\g(\vec{x} = \vec{x}_B, x^0 = 0)$ and the new term derives from
the existence of a subleading term $\beta(\vec{x}_B,0)$ in the expansion of $\phi$:
\be
\phi(X_B) \sim \beta(\vec{x}_B,x^0_B) z^{\Delta} + \g(\vec{x}_B,x^0_B) z^{d-\Delta} 
\ee
along the Euclidean section. For real times, under the quench, this expansion of $\phi(X_B)$ is modified
to \eqref{genexp2}, the new form is discussed in Appendix~\ref{app:ct} , where it is used to construct the general $S_{ct}$ quoted above. 
 Note that we have combined $S_{ct}$ with the CFT modular hamiltonian term  and subtracted the $\delta$ divergence found in Appendix~\ref{app:ct} 
and Appendix~\ref{app:divs1} from $\left< T_{\sigma\sigma} \right>_\g$ to define a ``renormalized'' CFT stress tensor. 
\be
 \left< T_{\sigma\sigma} \right>^{\rm ren}_\g  = \left< T_{\sigma\sigma} \right>_\g -  \left< T_{\sigma\sigma} \right>_\g^{\rm div}
  =  \left< T_{\sigma\sigma} \right>_\g - c' (2 \delta)^{d - 2\Delta} \g^2
\ee
After we have finished this computation we no longer
need any complicated contour prescription to compute $\phi$ on the Euclidean section.

To make closer contact with holography we now introduce a linearized metric $h$ about
the bulk of (Euclidean) $AdS_{d+1}$. We think about this 
metric as a book keeping device that will allow us to further package the calculation of EE into a single quantity, the area of this metric. As we will see there are certain consistency conditions which forces Einstein's equations to hold if we demand entropy is proportional to area. This packaging then
reduces the problem of calculating EE in any deformed CFT (at second order in perturbation theory), to that of a classical GR problem in one higher dimension.

We want to constrain $h$ so that the change in area due to $h$ of the RT surface $m(A)$ associated to the boundary region $A$
is related to the change in EE:
\be
\label{metric}
\delta S_{EE} = K \delta A(h)  \qquad \delta A(h) = \frac{1}{2R} \int_{ |\vec{x}|<R} d^{d-1} \vec{x}_B
z^{2-d} \left( \delta^{ij}R^2  -x^i_B x^j_B \right) h_{ij}
\ee
where $K$ is for now some unfixed constant. We will return to the problem of fixing $K$ later.
We have picked radial gauge for the metric fluctuations:
\be
h_{zz} = h_{z\mu} = 0 
\ee
and the area element in \eqref{metric} is integrated along $z^2 =  R^2- \vec{x}^2_B$. We have also,
for simplicity, chosen our entanglement  cut to be a sphere around the origin at $\vec{x}=0$.

For a general metric fluctuation $h$ we can now use a result derived in \cite{Faulkner:2013ica} using the machinery setup by Wald and Iyer \cite{Iyer:1994ys,Wald:1993nt}. We start by defining a $d-1$ form:
\be
\bm{\chi} \equiv - \frac{1}{2} \left( \delta ( \nabla^a \xi^b_B d \Sigma_{ab} )
+ \xi^b_B d \Sigma_{ab} \left( \nabla_c h^{ac} - \nabla^a h^c_c \right) \right)
\ee
where $d \Sigma^{ab}$ is the natural $d-1$ volume form for the metric $g = g^0 + h$. Explicitly:
\be
d  \Sigma^{ab} = \frac{1}{(d-1)!} \sqrt{-g} \epsilon^{ab}_{\,\,\,\,c_1 \ldots c_{d-1}} dx^{c_1} \wedge dx^{c_2 } 
\ldots \wedge d x^{c_{d-1}}
\ee
where our conventions are such that $\epsilon_{z\sigma x_1 x_2\ldots} = +1$.
Note that we only define $\bm{\chi}$ to first order in the $h$ fluctuations.

 A simple application
of Stoke's theorem for $d \bm{\chi}$ integrated over the region $A_B$ gives the following result:
\be
\label{waldiyer}
\delta A(h)  = - \int_{A_B} d \Sigma^a \xi_B^b \delta G_{a b} + 
\half \int_A d^{d-1} \vec{x}_B   \xi^\sigma z^{3- d}\left.  \left( \partial_z + \frac{2}{z} \right) 
\left(  \eta^{\mu\nu}  h_{\mu \nu}+ h_{\sigma\sigma} \right) \right|_{z = z_\Lambda}
\ee
where we have used the fact that $d \bm{\chi} = - d\Sigma^a \xi_B^b \delta G_{ab}$ is proportional to the linearized Einstein's equations without source:
\be
\delta G_{ab} \equiv \delta \left( R_{ab} - \half R g_{ab} - \half  d(d-1) g_{ab} \right)
\ee 
The area term defined in \eqref{metric} comes from $\int_{m(A)} \bm{\chi}$ by construction.
The last term in \eqref{waldiyer} comes from the boundary term $\int_A \bm{\chi}$ at $z = z_\Lambda$.

We we would like to compare  to the Wald-Iyer theorem \eqref{waldiyer} to our
results on perturbative calculations of EE. Combining \eqref{ee1}  with \eqref{ct2} we have
\be
\label{ee2}
\delta S_{EE} = -  2 \pi \int_{A_B} d \Sigma^a \xi_B^b T_{ab}^B+  2\pi \int_A d^{d-1} \vec{x}_B \xi^\sigma \left( - \Delta \beta \g - \frac{( d- \Delta)}{2} \g^2 z_\Lambda^{d-2 \Delta}
+ \left< T_{\sigma\sigma} \right>^{\rm ren}_\g \right) 
\ee
It is clear the various terms in \eqref{ee2}  and \eqref{waldiyer} can be identified after
using the proportionality of area and EE in \eqref{metric}. For example by considering
$\eqref{ee2} - K \times \eqref{waldiyer} = 0$. In particular by generalizing
the region $A$ via translations and rotations as well as considering
regions of all sizes $R$  we can arrive at the following statement -  
by demanding that the metric perturbation $h$ have the following boundary expansion:
\be
\label{bdryexp}
h_{\mu\nu} =\frac{4\pi}{d K} \left(  \left< T_{\mu\nu} \right>_\g^{\rm ren} -
\frac{\eta_{\mu\nu}}{d-1} \left(  \left< T^\mu_{\,\,\mu} \right>_\g^{\rm ren} + \Delta \beta \g \right) 
\right) z^{d-2} - \frac{\pi}{(d-1)K}  \eta_{\mu\nu}  \g^2 z^{2(d-\Delta-1)}  + \ldots
\ee
and that the metric perturbation satisfies the linearized Einstein equation in the bulk coupled
to the stress tensor of the field $\phi$:
\be
\label{einstein}
\delta G_{ab} = \frac{2\pi}{K} T_{ab}^B
\ee
then  \emph{the perturbed EE in the QFT can be calculated via the area entropy relation \eqref{metric}.}
This is then equivalent to the minimization procedure outlined in the introduction at
first order in the metric perturbation, due to the fact that the  RT surface $m(A)$
is a minimal area surface for the unperturbed metric, so to first order in the metric perturbation we
do not need to re-minimize the surface. 

Note that in order to complete the program of calculating EE, as stated, we also need to calculate $\left< T_{\mu\nu} \right>_\lambda$.
This can be done by integrating the CFT  three point function $T \mathcal{O} \mathcal{O}$ given in 
\eqref{too} against the couplings $\lambda(x)$ . We need this as an input to Eintein's equations since solving \eqref{einstein}
close to the boundary at $z=0$ with the assumption that $h_{\mu\nu} z^2 \rightarrow 0$  we can only reproduce the last term in \eqref{bdryexp}. The other terms are integration constants and so can only
 be fixed by some boundary condition at $z \rightarrow \infty$.
Thus it is natural to guess that a regularity condition should fix \eqref{bdryexp}. If this
is the case then actually we can invert this relationship to find $\left< T_{\mu\nu} \right>_\lambda$
in terms of the boundary expansion of $h$.  \emph{In fact this is the usual holographic prescription
for calculating the stress tensor} \cite{Balasubramanian:1999re}.\footnote{Comparing
to \cite{Balasubramanian:1999re} we find the same answer up to the terms with non-zero trace.  Such terms come about here
because of the relevant deformation. Note that the CFT stress tensor, $\left<T_{\mu\nu}\right>_\lambda$, in the presence
of these deformations is no longer traceless. Using the trace ward identity one can
show that $\left< T^\mu_{\,\,\mu} \right>^{\rm ren}_\lambda = -  \Delta (2\Delta -d) \beta \lambda$.
The terms proportional to $\eta_{\mu\nu}$ agree with the AdS/CFT results in \cite{Amsel:2006uf}, after quite a bit
of work to translate conventions. For example the correct QFT stress tensor in the presence
of this deformation is $\left< T^{\rm QFT}_{\mu\nu}\right>_\lambda = \left<T_{\mu\nu}\right>^{\rm ren}_\lambda + (2\Delta -d) \beta \lambda \eta_{\mu\nu}$ due to the additional $+\int \sqrt{g} \lambda \mathcal{O}$ term
in the action. We have used $\left<\mathcal{O}\right>_\lambda = (2\Delta - d) \beta$. }

So to summarize we would like to check that the boundary condition we quoted for $h_{\mu\nu}$
implies regularity as $z \rightarrow \infty$ for this metric perturbation.  We give the following argument - only for couplings $\lambda(x^0,\vec{x})$ which have a profile that die at large $|x| \rightarrow \infty$ sufficiently fast do we expect there to be some nice regularity condition in the bulk. In this case we
can take $A$ to be very large and the RT surface probes deep into the bulk. We can then use
the fact that $S_{EE}(A) = S_{EE}(A^c)$ for pure states (which should be the case
for such bounded couplings.) Now since in  the region $A^c$ the coupling goes to zero the perturbation to EE away from the CFT result also goes to zero and thus, via the entropy area relation, $h$
must vanish as $z \rightarrow \infty$. Of course many of the details, such as how fast the coupling
must vanish and how fast $h$ must vanish, have not been discussed. It would also be more satisfying to have an argument purely based on the linearized Einstein's equations. We leave this to future work.

One extension of this argument that may be attempted, is essentially the converse. For
example one might want to show that $h$ satisfying \eqref{einstein} and \eqref{bdryexp} is the \emph{unique} metric who's area encodes the EE of the QFT. We have shown that $h$ is one such metric.
This might be achieved following closely the arguments of \cite{Faulkner:2013ica}. We also leave this
to future work, although it is not clear to the author that it is important to establish such a statement.

Finally we would like to fix the constant $K$. For Einstein gravity we would have $K= 1/4G_N$.
However we have not identified this parameter in the field theory yet. In order to fix $K$
we should demand that the area entropy relation \eqref{metric} also works for the unperturbed metric
in the CFT.
The area of the unperturbed metric is then just related to the divergent volume of $\mathbb{H}_{d-1}$ space where we cutoff the volume integral  at large $Y^I = R/\epsilon$, for some UV cutoff $\epsilon \ll 1$. Keeping only the universal terms:
\be
S_{EE}(CFT) = K {\rm vol}(\mathbb{H}_{d-1} ) = K \frac{\pi^{h-1}}{\Gamma(h)} \left\{ \begin{matrix} \pi
(-1)^{h-\half} & \quad d \in {\rm odd}\\ - 2 (-1)^h \log R/\epsilon & \quad d \in {\rm even}
 \end{matrix} \right.
\ee
Comparing to \eqref{fanda} we have:
\be
K = \frac{\Gamma(h)}{\pi^h} \left\{ \begin{matrix}  (-1)^{h-\half}\log Z(S^d) & \quad d \in {\rm odd} \\
 \pi A_d  & \quad d \in {\rm even} \end{matrix} \right.
\ee
where these are quantities intrinsic to the odd (the sphere partition function $Z(S^d)$) or even (the Weyl trace anomaly coefficient $A_d$) dimensional CFT.

\section{Discussion}
\label{conclusions}

We have shown that the EE for deformed CFTs can be computed efficiently at second order in perturbation theory. The answer lends itself to a holographic interpretation in terms
of an emergent higher dimensional gravitational theory. It is not surprising that EE is the correct observable to talk about the emergence of gravity in QFTs. We just need to develop
better tools to calculate EE to realize this fully. In this paper we have made some initial steps. We end with some discussion of the meaning and context of this result in AdS/CFT
as well as mentioning some further work.

\subsection{Expectation for gravitational emergence}

Since the gravitational interpretation of EE established here works for any deformed CFT, one might
ask how the usual expectations for gravitational emergence in AdS/CFT fit into this story.
The answer to this questions lies in the fact that we are working in perturbation theory,
which roughly speaking fixes us close to the $AdS$ boundary. Here the physics is somewhat
universal since fluctuations die off close to the boundary.  Moving to higher orders in perturbation theory we should see the classical gravitational description breakdown - both quantum
effects \cite{Barrella:2013wja,Faulkner:2013ana,Engelhardt:2014gca} and higher derivative corrections should be expected \cite{Dong:2013qoa,Camps:2013zua,Bhattacharyya:2014yga,Miao:2014nxa}.
That is, of course, unless we are working with a special CFT to begin with - one that we might have expected to have a classical gravitational description, for example via the conditions
discussed in \cite{Heemskerk:2009pn,ElShowk:2011ag}. 

If this is the case it would be interesting to extend these calculations to see the various
 predictions for the behavior of EE in holographic theories. For example
one of the usual requirements for bulk emergence is that the CFT allows for a large-$N$ limit in which correlation functions of certain special single trace operators factorize. This then corresponds
to a classical limit for the bulk theory. In this limit one may
expect to have saddle type behavior for EE from which interesting phase transitions 
can ensue \cite{Headrick:2010zt}.  Of course in perturbation theory we cannot expect to see such behavior.  
One clearly needs to sum an infinite number of terms in perturbation theory.

\subsection{Real times}

The results we quote are for calculating EE in the presence of an in-homogenous deformation in space and in imaginary time. It is natural to ask how the calculation changes when we have a time
dependent coupling in real times. For this we need to develop the replica trick in real times and
presumably this will have some bearing on the HRT conjecture \cite{Hubeny:2007xt} for how to generalize Ryu-Takayanagi to real times. We could also hope to find universal terms in the time dependence
of EE after a quench close to a CFT fixed point, extending the interesting results in \cite{Buchel:2013gba,Das:2014hqa}. 

We should also mention here the peculiar real time prescription that
was forced upon us when we did this calculation. We found that we should
set the coupling $\g$ to zero in real times when computing the integral of the bulk modular
hamiltonian over some surface which extends into real times. It would be good to find an interpretation
of this in holography. 
In some sense this is the only prescription we could have expected
since we did not specify the behavior of the real time couplings - so setting the coupling
to zero is the only universal procedure one can think of. It may be possible there
is a way to avoid this prescription when we do have a specified time dependent coupling in real times.

One extension along these lines relates to studying the state dependence of EE when
we perturb away from the ground state. One can show there is an intimate
relation between the gravitational dynamics in holographic theories and the first law of entanglement for small perturbations of the CFT vacuum state \cite{Blanco:2013joa,Lashkari:2013koa,Faulkner:2013ica,Swingle:2014uza}.\footnote{These results mimic and
were inspired by the insights of Jacobson \cite{Jacobson:1995ab} on the relation between Einstein equations
and thermodynamics. Interestingly Jacobson
works with boost energy integrated across a rindler horizon, so there may be a more direct
connection with our results in their initial form around \eqref{bulkmod} before we pushed them away from
$\mathcal{H}^+$ to the space like surface $A_B$. }
Such states can be specified using a euclidean path integral in the presence of coupling deformation for various operators.  Moving into real times, after setting these couplings to zero, results
in a non-trivial state in the undeformed theory.  Along these lines we could use the method developed in this paper to compute, for example, the relative entropy between the excited state and the vacuum state and compare to the results from holography as in \cite{Banerjee:2014oaa,Lin:2014hva,Lashkari:2014kda,Bhattacharya:2014vja}.

\acknowledgments

We would like to thank  Srivatsan Balakrishnan, Souvik Dutta, Jared Kaplan, Rob Leigh, Juan Maldacena, Tassos Pekou for useful discussion and comments.  TF acknowledges support from the UIUC Physics Department.

\appendix

\section{More on embedding coordinates}
\label{app:maps}

In this appendix we setup some notation relating to the embedding space coordinates $P$. For a nice review of this formalism see \cite{SimmonsDuffin:2012uy}.  For the definition of $P$ see Section~\ref{setup} around
\eqref{defemb}.

At various points in this paper we would like to define integration over $P$ and it is sometimes useful to think of these integrals as conformal
integrals. Such integrals were crucial for the recently developed methods to compute conformal blocks in CFTs \cite{SimmonsDuffin:2012uy}. Following that paper
one can define integrals over the projective coordinates as follows:
\be
\int d P f(P) \equiv \int \frac{ d^{d+2} P}{{\rm Vol} \mathcal{G}^+} \delta ( P^2 ) \theta(P^I)  f(P)
\ee
where we have fixed to the future light cone of $P$ and divided by the volume of the gauge group related to projective rescalings $P \rightarrow \Lambda P$. So here $\mathcal{G}^+ = GL(1)^+$ is the (connected) group of boosts in $1+1$ dimensions.  
This only works if $f(P)$ has the correct weight under the rescaling.
 That is 
$f( \Lambda P) = \Lambda^{-d} f( P)$. 
While many of the integrals we consider naively don't have this property, due to the fact
that we are breaking conformal invariance by the $\g$ deformation, we can fix this
scaling by multiplying by factors of $P \cdot P_\infty$ where $P_\infty$ is a spurious coordinate which we define to be the point at infinity for the flat space $\mathbb{R}^d$ gauge:
\be
P_\infty \equiv \frac{1}{2R} \left( 1 , -1 , 0, \ldots \right)
\ee

In flat coordinates we have simply:
\be
\int d P f(P)= \int d^{d} \vec{x}f(P \rightarrow P_F)
\ee
and in Hyperbolic slicings we have:
\be
\int d P f(P) = \int_0^{2\pi} d\tau \int d Y f(P \rightarrow P_H)
\ee
where we have defined integration over the Hyperbolic coordinates $Y$ as:
\be
d Y \equiv d^{d}Y \delta(Y^2+1)  \theta(Y^I)
\ee
where $Y = (Y^I, Y^m)$ for $m=1 \ldots d-1$ such that:
\be
\mathbb{H}^{d-1} \, : \, \, Y^2 = - (Y^I)^2 + Y^m Y^m =  -1 \qquad Y^I >0
\ee
Finally the distance functions on these spaces can be defined via the natural product:
\be
P_{12} = - 2 P_1 \cdot P_2 \qquad Y_{12} = -2 Y_1 \cdot Y_2
\ee
For projective space these distances depend on the gauge choice, which becomes
the statement that CFT correlation functions pick up conformal factors under conformal maps:
\be
\label{distconf}
(x_1 - x_2)^2 = \left. P_{12} \right|_F = \Omega_1^{-1} \Omega_2^{-1} \left( - 2 \cos(\tau_1 - \tau_2)
- 2 Y_1 \cdot Y_2 \right)
\ee
where the geodesic distance $d(1,2)$ between the two points on $\mathbb{H}_{d-1}$ satisfies $\cosh d(1,2) = - Y_1 \cdot Y_2$.  The conformal factors above are defined in \eqref{conff}

\section{Continuation in $n$}
\label{app:carlson}

For finite quantum systems it is clear, since $\tr \rho_A^n$ has well defined
analytic properties in $n$. In particular Carlson's theorem applies \cite{Carlsons} and we can use this to define an analytic continuation away from the integers. That is given a function $f(z)$  that we know at integer values $z = n$ for $n=1, \ldots \infty$, and additionally assuming that the function is \emph{analytic}
 and exponentially bounded for  ${\rm Re}(z)>1$ and dies more rapidly than $\sin(\pi z)$ as $z \rightarrow 1 \pm  i \infty$ we can uniquely construct $f(z)$ from this data.  

For us the situation is less clear, since for a QFT even the integer Renyi entropies are not
well defined - they are UV divergent. After we subtract the appropriate divergences
their analytic behavior in $n$ is far from clear. For example if we regularize the QFT on the lattice
so that we have a finite quantum system, then it is not clear the continuum limit
commutes with statements about the analytic properties of the subtracted version of $\tr \rho_A^n$. 
Indeed there are well known examples where this does not happen. For example the various
phase transitions as a function of $n$ that occur when calculating the Renyi entropies \cite{Belin:2013dva,max}. In these cases $n$ can be usefully thought of as an inverse temperature, and so it should come
as no surprise that a phase transition signals a non-analyticity in the complex $n$ plane.

Presumably this non-analyticity is not an essential aspect of the necessary continuation
of integer Renyi entropies in order to calculate EE, in particular it usually occurs well away
from $n \approx 1$.  We will take the viewpoint that we should
remove as many of the non-analyticities in the complex $n$ plane as we can, in order
to define the correct continuation. 

\section{Direct Calculation}
\label{app:direct}

Start with the identity:
\be
- \ln \rho = \int_0^\infty d\beta \left( \frac{1}{\beta + \rho} - \frac{1}{\beta+1} \right)
\ee
where $\rho = \rho_A$ is the reduced density matrix for region $A$. So the EE can be written:
\be
S_{EE} = - {\rm tr} \rho \ln \rho = \int_0^\infty d \beta \left( {\rm tr}\left( \frac{\rho}{\beta + \rho} \right)-   \frac{1}{\beta+1} \right)
\ee
The term we are interested in, which we call $\delta S_{EE}^{(2)}$ in the main text, comes from the second order variation in $S_{EE}$ due to a first order change in $\rho = \rho_0 + \delta \rho$.
The contribution from the second order change in $\delta \rho$ gives $\delta S_{EE}^{(1)}$. This
term is more straightforward to deal with so we don't consider it in this appendix.  We have:
\be
\delta S_{EE}^{(2)} = - \int_0^\infty d \beta \beta \, \, {\rm tr} \left( \frac{1}{(\beta + \rho_0)^2} 
\delta \rho \frac{1}{(\beta + \rho_0) } \delta \rho \right) 
\ee
We now write the perturbation of the density matrix as:
\be
\delta \rho = \g \int_{\mathbb{H}_{d-1}} dY \int_0^{2\pi} d \tau \rho_0 U( \tau) \widehat{\mathcal{O}}(0,Y) U(-\tau) \Omega^{\Delta-d} (\tau,Y)
\ee
where $U(\tau) = \rho_0^{ \tau/(2\pi)}$. Putting everything together:
\be
\delta S_{EE}^{(2)} = -2 \int d\mu
\int_0^\infty d \beta \beta \, \, {\rm tr} \left( \frac{\rho_0}{(\beta + \rho_0)^2} 
\widehat{\mathcal{O}}(i \tau_a,Y_a) \frac{\rho_0}{(\beta + \rho_0) } \widehat{\mathcal{O}}(i \tau_b,Y_b) \right) 
\ee
where we use the notation introduced in \eqref{dmu} for integrating over the two positions. 
We now insert a complete set of states labeled by the energies or entanglement eigenvalues.
This gives the form of a (finite temperature) spectral representation:
\be
\delta S^{(2)}  = -2 \int d\mu \int_{\omega} \int_{\omega'}
| \left< \omega \right| \widehat{\mathcal{O}}\left| \omega' \right> |^2 e^{ - (\omega - \omega')(\tau_a -\tau_b)}
\int_0^\infty d \beta \beta \frac{ e^{-2\pi(\omega+\omega')}}{ (\beta + e^{-2\pi \omega})^2 (\beta + e^{-2\pi \omega'})}
\ee
This last integral can be done and we have:
\be
\delta S^{(2)}  = -2 \int d\mu \int_{\omega} \int_{\omega'}
| \left< \omega \right| \widehat{\mathcal{O}} \left| \omega' \right> |^2 e^{ - \nu(\tau_a -\tau_b)/2\pi} e^{ -2\pi \omega'}
\left( \frac{1}{1 - e^{\nu}} + \frac{\nu e^{\nu}}{(1 - e^\nu)^2} \right) 
\ee
where $\nu =2\pi( \omega- \omega')$.  Now we can write:
\be
\left( \frac{1}{1 - e^{\nu}} + \frac{\nu e^{\nu}}{(1 - e^\nu)^2} \right)  =
\int_{-\infty- i\epsilon}^{\infty - i \epsilon} \frac{d s}{2 \pi i} e^{ - i \nu s/2\pi}   \frac{s}{4 \sinh^2(s/2)}
\ee
and then undo the spectral representation of the two point function. This procedure will  only work for
$\tau_b < \tau_a$ for convergence reasons. Similarly we can write:
\be
\left( \frac{e^\nu}{1 - e^{\nu}} + \frac{\nu e^{2\nu}}{(1 - e^\nu)^2} \right)  =
\int_{-\infty+ i\epsilon}^{\infty + i \epsilon} \frac{d s}{2 \pi i} e^{ - i \nu s/2\pi}   \frac{s - 2 \pi i }{4 \sinh^2(s/2)}
\ee
This will work for $\tau_a < \tau_b$. To organize this properly we split the $\tau$ integrals into 
time-ordered segments:
\be
\delta S^{(2)}(\tau_b<\tau_a) = - 2\int_{\tau_b < \tau_a} \hspace{-.3cm} d\mu\int_{- i \epsilon} \frac{ ds}{ 2\pi i} \frac{s}{4 \sinh^2(s/2)}
 {\rm tr} \left( \rho_0 \widehat{\mathcal{O}}(i \tau_b + s) \widehat{\mathcal{O}}(i \tau_a) \right)
\ee
and
\be
\delta S^{(2)}(\tau_a<\tau_b)  = - 2\int_{\tau_a < \tau_b} \hspace{-.3cm}  d\mu \int_{+ i \epsilon} \frac{ ds}{ 2\pi i} \frac{s- 2 \pi i}{4 \sinh^2(s/2)}
 {\rm tr} \left( \rho_0\widehat{\mathcal{O}}(i \tau_a)  \widehat{\mathcal{O}}(i \tau_b + s )  \right)
\ee
On this last expression we can now relabel the $\tau_a \leftrightarrow \tau_b$ integrals as well as the spatial integrals $Y_a \leftrightarrow Y_b$.
If we also relabel $s \rightarrow -s$ we derive the relationship:
\be
\delta S^{(2)}(\tau_a<\tau_b) + \delta S^{(2)}(\tau_b<\tau_a) = 2 \int_{\tau_b < \tau_a} d\mu \int_{- i \epsilon}\frac{ds}{4 \sinh^2(s/2)}
 {\rm tr} \left( \rho_0\widehat{\mathcal{O}}(i\tau_a)  \widehat{\mathcal{O}}(i\tau_b + s )  \right)
\ee
Adding this equation to the same equation with coordinates switched (and $s \rightarrow -s$) we find:
\be
\delta S^{(2)}  = \int d\mu\int_{C(\tau_{ab})} \frac{ds}{4 \sinh^2(s/2)}
 {\rm tr} \left( \rho_0 \mathcal{T} \widehat{\mathcal{O}}(i \tau_a)  \widehat{\mathcal{O}}(i \tau_b +  s )  \right)
\ee
where we have a time ordered correlator and the $s$ integration contour 
depends on the ordering of $\tau_a$ and $\tau_b$. That is $C = s + i \epsilon\, {\rm sgn} (\tau_a - \tau_b)$.
This is the final expression with the correct contour prescription. A simple contour deformation
 then lands us on the answer given in \eqref{tord} which was derived using the replica method.

\section{Divergence in $\delta S^{(1)}$}
\label{app:divs1}

Here we are interested in finding the divergent term in \eqref{s1h}. Recall that we regulate
this  divergence by  only integrating over the region $\mathcal{C}(\delta) : |x^0_{a,b}| > \delta$. We will work with general spatial dependent couplings.
We need to calculate: \be
\left< T_{\sigma\sigma} (x^0 = 0,\vec{x}) \right>_\g \equiv - \frac{1}{2} \int_{ \mathcal{C}(\delta)} d^{d} x_a   \int_{ \mathcal{C}(\delta)} d^{d} x_b 
\left< T_{00} (0,\vec{x}) \mathcal{O}(x_a) \mathcal{O}(x_b) \right> \g(x_a) \g(x_b)
\ee
where we wick rotate to imaginary times $\sigma = i x^0$.
By examining the following form (which we really won't need) of the $T\mathcal{O}\mathcal{O}$
three point function we can see that for $\Delta < d$ the only  divergence in the above integrals comes from when $x_a \rightarrow x_b \rightarrow x$ (we have set $x=0$ to save space):
\be
\label{too}
\left< T_{00} (0) \mathcal{O}(x_a) \mathcal{O}(x_b) \right> =
 \frac{c^T_\Delta}{ |x_{ab}|^{2\Delta - d +2}|x_{a}|^d |x_{b}|^d}
 \left( \left( x_a^0 \frac{|x_{b}|}{ |x_{a}|}-x_b^0 \frac{|x_{a}|}{ |x_{b}|} \right)^2 - \frac{1}{d} |x_{ab}|^2
 \right)
\ee
where $c^T_\Delta= d c_\Delta \Delta/(d-1)/S_{d-1}$  with $S_{d-1}$ the area of
a $d-1$ sphere and where $c_\Delta$ sets the normalization of the CFT 2 point function.
That is we can expand $\g(x_{a,b})$ around $\g(x)$ and proceed with the integrals:
\be
\label{Tdiv}
\left< T_{\sigma\sigma} (0,\vec{x}) \right>_\g^{div}
 \sim \g(0,\vec{x})^2  J \, \qquad J =- \frac{1}{2} 
  \int_{ C(\delta)} d^{d} x_a   \int_{ C(\delta)} d^{d} x_b  
\left< T_{00} (0,\vec{x}) \mathcal{O}(x_a) \mathcal{O}(x_b) \right> 
\ee
after which the IR bounds on the integrals in $J$ can be pushed to infinity since there is no IR divergence to speak of for $\Delta > d/2$.
We can calculate $J$ as follows. Firstly note that since $\delta$ is the only scale
in the above integral the answer must be proportional to $\delta^{-2\Delta+d}$ which
diverges for $\Delta > d/2$. Since additionally the answer for the integral is independent of $\vec{x}$
we can integrate $J$ over $\vec{x}$ and divide by the spatial volume. This integrates
to a conserved charge (the energy associated to $\partial_0$)
\be
J = - \frac{1}{2V}  \int_{ C(\delta)} d^{d} x_a   \int_{ C(\delta)} d^{d} x_b   \int d^{d-1}\vec{x} 
n^\mu \left< T_{0\mu} (0,\vec{x}) \mathcal{O}(x_a) \mathcal{O}(x_b) \right> 
\ee
where $n = \partial_0$ is normal to the surface defined by the $\vec{x}$ integral.
We can now deform this $x$ surface integral, as long
as we stay away from the operator insertions. If both operators are on the same
side of the region $|x_0| > \delta$ then there is no divergence and the answer is
clearly zero. However if the operators are on different sides then there can be a 
pinching divergence where the two operators come towards $x^0 = 0$ from
opposite sides. 

We can find this divergence by deforming the $\vec{x}$ integral so it encircles (as a $S^{d-1}$ sphere) one
of the operators (say the $b$ operator), and then pushing the remaining part of the $x$ integral off to infinity where
it gives zero. The integral encircling the operator simply gives $-\partial_0 \mathcal{O}$, thanks
to the energy Ward identity and we are left with: 
\begin{align}
J & = \int_{x^0_a > \delta} d x_a^0 \int_{x^0_b < - \delta} d x_b^0
\int d^{d-1} \vec{y}  \frac{\partial}{\partial x_b^0} \left< \mathcal{O}(x^0_a,0) \mathcal{O}( x^0_b,
\vec{y}) \right> \\
&= \int_{\delta}^\infty d x^0_a \int d^{d-1} \vec{y} \frac{  c_\Delta}
{\left( (\delta + x_a^0)^2 + \vec{y}^2 \right)^\Delta} = \frac{ \Gamma(\half - h + \Delta)}{ \sqrt{\pi}  \Gamma(-h + \Delta)} (2 \delta)^{ d- 2 \Delta} \equiv  c' (2 \delta)^{ d- 2 \Delta} 
\label{cpdef}
\end{align}
where this last line defines $c'$.
Integrating the divergent term in \eqref{Tdiv} over $2\pi \xi^0$ gives the claimed expression
in \eqref{ddiv}.

\section{Counter terms}
\label{app:ct}

In this appendix we will carefully construct the counter terms quoted in \eqref{ct1} and \eqref{ct2}.
These come from the ``vertical'' part of the integral over the bulk stress tensor when we
push away from the horizon $\mathcal{H}^+$. We start
with the general expression for $\phi(X)$ with the appropriate real time prescription discussed
in the body of the paper. In this appendix we work with a non constant coupling $\g(x)$. That is:
\be
\label{appint}
\phi(X) = \frac{c_\Delta}{2 (\Delta -h)} \int d^{d-1} \vec{x} \int_{C(\delta)} d x^0
\frac{ \g(x^0,\vec{x}) z^\Delta}{ \left( z^2 + (\vec{x} - \vec{x}_B)^2 + (x^0 - i \sigma)^2 \right)^\Delta}
 \ee
We now take a limit of this expression close to the boundary $z = z_\Lambda \rightarrow 0$ as well
as small $\sigma \sim z_\Lambda$. As we do this
there is a singular part coming from $\vec{x} \approx \vec{x}_B$ and for $ x^0 \approx \sigma \approx 0$. On the Euclidean section this evaluates to a delta function, however in real times we find
a more general answer:
\be
\label{genexp2}
\phi(X) \approx \beta(0,\vec{x}_B) z_\Lambda^\Delta + \g(0,\vec{x}_B)
z_\Lambda^{ d-\Delta} F_\delta( \sigma/z_\Lambda) 
\ee
where $F_\delta$ is the same function we found for the uniform coupling case \eqref{Ffirst}. 
The first term in \eqref{genexp2} above is just the smooth part of the integral \eqref{appint}.
As was the case for the uniform coupling, there can be no finite contribution to the EE
from the region where $\sigma \ll z_\Lambda$ since real time coupling is turned off and the
sub-leading term can not contribute a finite amount to this ``vertical'' integral. 
Thus we concentrate on the region $0< \sigma \lesssim z_\Lambda$ and take a scaling limit
where the only term surviving in the bulk stress tensor is:
\begin{gather}
S_{ct} =  2\pi \frac{1}{z_\Lambda^{d-1}}  \int_{\mathcal{D}^+(A)}d\sigma d^{d-1} \vec{x}_B
\xi^\sigma \partial_z \phi \partial_\sigma \phi \\  =
2\pi \int_{\mathcal{D}^+(A)} d\sigma d^{d-1} \vec{x}_B \xi^\sigma \left( \Delta \beta \g \partial_\sigma F_\delta + (d-\Delta) z_\Lambda^{d-2\Delta}
F_\delta \partial_\sigma F_\delta  - z_\Lambda^{d-2\Delta} \sigma (\partial_\sigma F_\delta)^2 \right)
\end{gather}
where $\beta$ and $\g$ are to be evaluated at $(x^0 = 0,\vec{x}_B)$.
The first two terms can be integrated in $\sigma$ easily using $F_\delta(0) = 1$ and
$F_\delta(\infty) \rightarrow 0$ for $\Delta > d/2$ and these
give rise to the first two terms on the right hand side of \eqref{ct2}. The last term is a little tricky and actually gives a divergence in $\delta$. We take $\delta \ll z_\Lambda$. To evaluate this we take a further scaling limit such that $(\sigma - z_\Lambda)
= \hat{\sigma} \delta  $ as $\delta \rightarrow 0$. After doing this we effectively split the $\sigma$ integral into three parts matching at $\sigma_{1,2}$:
\begin{align}
\label{three}
\int & d\sigma \sigma (\partial_\sigma F_\delta)^2 =   - \half (c')^2 (2\delta/z_\Lambda)^{-2\Delta +d}  \int_{\hat{\sigma}_1}^{\hat{\sigma}_2}  d\hat{\sigma}
\left( (\hat{\sigma} + i)^{-\Delta + h - 1/2} -(\hat{\sigma} - i)^{-\Delta + h - 1/2}  \right)^2
 \\ \nonumber
 &  + \int_0^{z_\Lambda + \delta \hat{\sigma}_1} d\sigma  \mathcal{O}(\delta) 
  + 4 c'^2  \sin^2( \pi(\Delta - h+1/2)) (z_\Lambda)^{4\Delta -2 d} \int^{\infty}_{z_\Lambda + \delta \hat{\sigma}_2} d \sigma \sigma ( \sigma^2- z_\Lambda^2 )^{-2\Delta + (d-1)} 
\end{align}
where we take the matching points to satisfy $- \hat{\sigma}_{1}\gg 1 $ and $\hat{\sigma}_2 \gg 1$.
Note the branch cut prescription for the first line in \eqref{three} can be gleaned from
the integral prescription defining $F_\delta$, that is $C(\delta)$ shown in the right panel of
Figure~\ref{fig:zplane}. The second term in \eqref{three} is $\mathcal{O}(\delta)$ and so can
be ignored in this limit. The third term can be evaluated an gives rise to:
\be
4 c'^2  \sin^2( \pi(\Delta - h+1/2)) (2  \hat{\sigma}_2 \delta /z_\Lambda )^{ -2 \Delta + d} 
\ee
Note the upper limit on this last $\sigma$ integral has been extended to $\infty$ which works for $z_\Lambda \ll 1$ where we are focusing on the region close to $\sigma = 0$.
This last term is small for large $\hat{\sigma}_2$ and so we can ignore it. 
We are left with the first term in \eqref{three} which for $\hat{\sigma}_1 \rightarrow -\infty$
and $\hat{\sigma}_2 \rightarrow \infty$ gives only a non-zero contribution from the cross term in the
square:
\be
=  c'^2 (2 \delta/z_\Lambda)^{-2\Delta +d } \int_{-\infty}^\infty d\hat{\sigma} (\hat{\sigma}^2 + 1)^{-\Delta + h -1/2}
= c' (2 \delta/z_\Lambda)^{-2 \Delta + d}
\ee 
After integrating this over $-2 \pi z_\Lambda^{d-2\Delta} \int_A \xi^0$ 
this then gives rise to the last divergent term quoted in \eqref{ct1}, which cancels
the term discussed in Appendix~\ref{app:divs1}.

\end{document}